\documentclass[a4paper,fleqn]{cas-sc}

\usepackage[authoryear]{natbib}
\usepackage{hyperref}
\usepackage{graphicx}
\usepackage{pgffor}

\usepackage{longtable}
\usepackage{tabu}
\usepackage{multirow}
\usepackage{amsmath}
\usepackage{verbatim}
\usepackage{pdflscape}
\usepackage{caption}
\usepackage{subcaption}
\usepackage{amsmath}
\usepackage[normalem]{ulem}
\usepackage{gensymb}
\usepackage{adjustbox}
\usepackage[switch]{lineno} % Adds line numbers
\usepackage{soul}
\usepackage{float}
\usepackage{placeins}
\usepackage[switch]{lineno}
\usepackage{soul}
\usepackage[normalem]{ulem}

\DeclareCaptionFormat{cont}{#1 (cont.)#2#3\par}

\begin{document}
\let\WriteBookmarks\relax
\def\floatpagepagefraction{1}
\def\textpagefraction{.001}
\shorttitle{}
\shortauthors{D. Oszkiewicz et~al.}

\title [mode = title]{Asteroid phase curve modeling with empirical correction for shape and viewing geometry}                      

%\linenumbers

\tnotetext[1]{This document is the results of the research
   project funded by the National Center of Science, Poland.}

\author[1]{Dagmara Oszkiewicz}[orcid=0000-0002-5356-6433]
\author[1,2]{Przemysław Bartczak}[orcid=0000-0002-3466-3190]
\author[1]{Milagros Colazo}[orcid=0000-0001-6082-2477]
\author[3]{Antti Penttil\"{a}}[orcid=0000-0001-7403-1721]

\address[1]{Astronomical Observatory Institute, Faculty of Physics and Astronomy, A. Mickiewicz University, S{\l}oneczna 36, 60-286 Pozna{\'n}, Poland}
\address[2]{Instituto Universitario de F{\'i}sica Aplicada a las Ciencias y las Tecnolog{\'i}as (IUFACyT).
   Universidad de Alicante, Ctra. San Vicente del Raspeig s/n. 03690 San Vicente del Raspeig, Alicante, Spain
 \address[3]{  
   Department of Physics, PO Box 64, FI-00014 University of Helsinki, Finland}
   }% \address[2]{Instituto de Astrof\'isica de Andaluc\'ia, CSIC, Apt 3004, E18080 Granada, Spain}

% Activate line numbers

\begin{abstract}
We present a novel empirical method for correcting asteroid phase curves for rotational and geometrical effects using precomputed spin-and-shape models. Our approach normalizes sparse photometric data to a pole-on geometry, enabling consistent phase-curve fitting across apparitions. We fit both the $H,\!G_1,\!G_2$ and $H,\!G_{12}$ phase functions to the normalized data. We also numerically derive new constraints on parameter ranges that ensure physically meaningful solutions. These constraints are based on the requirement that the reduced magnitude must monotonically decrease with phase angle and remain within plausible slope bounds. Compared to earlier bounds, our new constraints are more permissive. We also compare derivative-based and derivative-free optimization methods, highlighting convergence issues with the $H,\!G_{12}$ function and offering mitigation strategies. We applied our method to over 25,000 asteroids observed by the ATLAS survey, demonstrating its usuability. The new method enables the selection of the preferred spin-and-shape solution based on either statistical phase-curve model selection criteria and/or physically motivated constraints on the phase-curve shape.
\end{abstract}

\begin{highlights}
\item We present a novel phase curve computation method accouting for shape and geometrical effects
\item We derive new boudnding conditions for phase curve parameters
\item We apply the new method and the new conditions to a sample of objects
\end{highlights}

\begin{keywords}
Asteroids
\end{keywords}

\maketitle

\section{Introduction}

Phase curves - describing the variation in an asteroid’s brightness as a function of the solar phase angle- are fundamental for characterizing the physical and surface properties of small Solar System bodies. They provide insights into surface composition, albedo, roughness, and regolith properties, and are essential for determining absolute magnitudes as well as deriving sizes and albedos, particularly when combined with thermal infrared data. Traditionally, phase-curve studies required time-consuming, dedicated observations spanning a wide range of phase angles and viewing geometries, thereby limiting the number of objects that could be studied. However, the growing availability of sparse photometric data from large-scale sky surveys enabled statistical analyses of phase curve behavior across hundreds of thousands of objects. This shift has driven the development of new methodologies capable of extracting reliable phase-curve parameters from heterogeneous sparse datasets sometimes calibrated in different ways.

The application of phase functions to large sets of sparse photometric data (typically originating from a single source) began over a decade ago. \citet{oszkiewicz2011online} derived phase curves for nearly 500,000 asteroids using sparse photometry from the Minor Planet Center (MPC). The primary challenge associated with this dataset was the integration of observations from various observatories and observers, taken at different times and characterized by differing and often unknown photometric uncertainties. Since then, substantial progress has been made both in terms of methodological advancements for handling sparse data and improvements in the quantity, quality, and integration of multiple data sources.

%Single data sources
Most early studies focused on homogeneous datasets from a single survey. For instance, \citet{verevs2015absolute} utilized uniform Pan-STARRS photometry to derive phase-curve parameters for approximately 300,000 asteroids. Similarly, \citet{waszczak2015asteroid} analyzed data from the Palomar Transient Factory (PTF) for around 50,000 objects, employing a combined rotation and phase-function model. More recently, \citet{mahlke2021} demonstrated a wavelength dependency in phase-curve parameters for 95,000 asteroids using sparse dual-band photometry from the ATLAS survey—a result independently confirmed by \citet{vdurech2020asteroid}, who analyzed the same dataset. \citet{colazo2025asteroid} extended this work by deriving phase curves for about 300,000 asteroids using the next ATLAS calalogue and data aggregated across multiple apparitions. In addition, \citet{alvarez2022phase} applied a Bayesian framework to the Sloan Digital Sky Survey (SDSS) data to extract 15,000 phase-curve parameters in different different filters and studied the phase coloring effect \citep{alvarez2024multiwavelength}.
% Combining different data sets
Some efforts have also focused on the integration of multiple photometric datasets. For example, \citet{colazo2021determination} and \citet{wilawer2022asteroid} combined Gaia DR2 photometry with ground-based observations, enhancing phase-curve derivation.

%Different methods for corretcing for the viewing and observing geometry 
Advancements have also been made in methods accounting for viewing and observational geometry. \citet{muinonen2020} introduced a Bayesian inversion technique capable of simultaneously determining rotation periods, pole orientations, convex shapes, and phase-curve parameters. Initially applied to three asteroids with extensive Gaia DR2 data, the method was subsequently scaled up by \citet{martikainen2021asteroid} to a sample of approximately 500 objects, and by \citet{wilawer2024} to 35 well-observed asteroids with both dense ground-based and sparse ATLAS photometry.
\citet{carry2024combined} proposed a method for jointly determining spin orientation and phase-function parameters, incorporating a geometric model to account for brightness variations induced by spin-axis orientation and polar oblateness. This approach was applied to nearly 100,000 asteroids observed by the Zwicky Transient Survey (ZTF) and has been integrated into the FINK broker system \citep{moller2021fink}. 

The current method represents yet another methodological approach, an intermediate step between full spin, shape, and phase-curve inversion and the geometric model proposed by \citet{carry2024combined}. It provides a valuable alternative, balancing complexity and computational efficiency while enabling the use of large, sparsely sampled datasets. In Section \ref{met} we describe the methodology, the results are in Sec. \ref{res} and the conclusions in Sec. \ref{concl}.

\section{Methods}
\label{met}

\subsection{Normalization of observed magnitudes to remove rotational and aspect-induced variations}

Generally, the apparent visual magnitude $V$ of an asteroid can be modelled as:

\begin{equation}
V = 5 \log_{10}(r \Delta) - \Phi (\alpha, H, G_1, G_2) + \delta,
\end{equation}

where $H$ is the absolute magnitude, which generaly may vary between oppositions due to changes in aspect and viewing geometry, however here denotes the absolute magnitude corresponding to the pole-on geometry; $\delta$ represents the brightness modulation caused by the asteroid’s rotation and irregular shape (described by a large number of free parameters—typically several tens to hundreds) and due to geometry/aspect; $r$ and $\Delta$ are the heliocentric and geocentric distances, respectively (in AU); and $\Phi(\alpha, H, G_1, G_2)$ is the phase function, which depends on the phase angle $\alpha$. In particular, we adopt the $H$, $G_1$, $G_2$ phase function \citep{muinonen2010}, as recommended by the IAU since 2012. However any other phase function can also be utilized.

\begin{figure}[ht]
    \centering
    \includegraphics[width=0.45\linewidth]{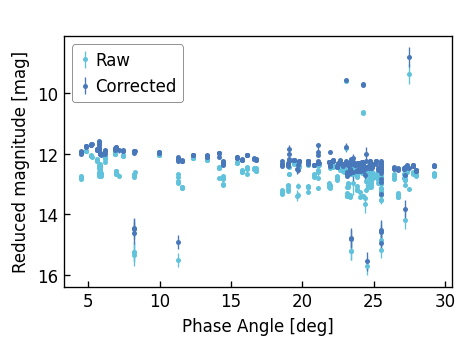}
    \includegraphics[width=0.45\linewidth]{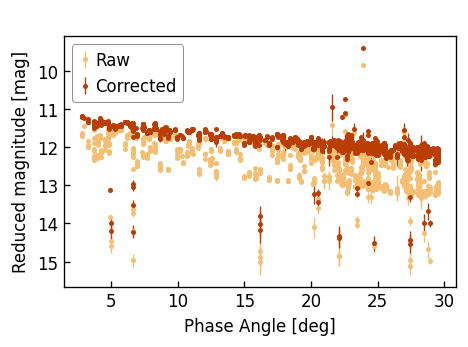}     \\
    \includegraphics[width=0.45\linewidth]{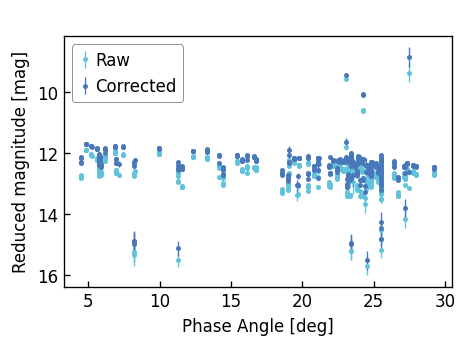}
    \includegraphics[width=0.45\linewidth]{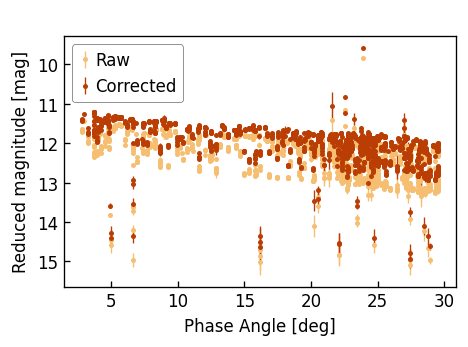}     \\
    \includegraphics[width=0.45\linewidth]{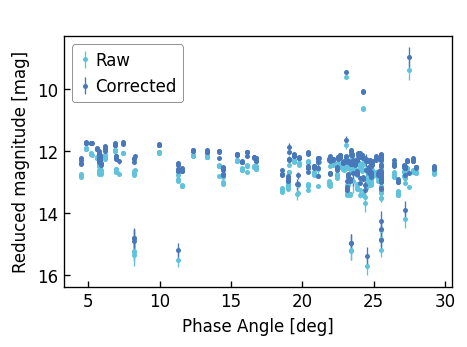}
    \includegraphics[width=0.45\linewidth]{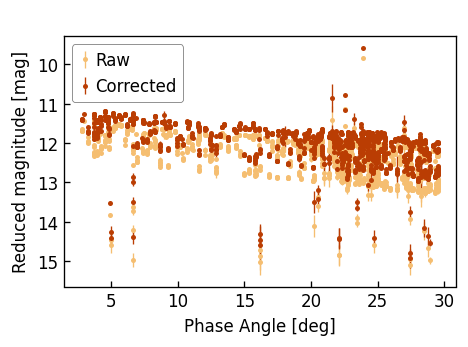}     \\
   \caption{Data normalization for asteroid (1182) Ilona. Raw and geometry-corrected photometric data are denoted in colors. The left panels correspond to the ATLAS c filter and the right panels to the o filter. Rows show results for the following shape models: Cellinoid (top), DAMIT model 1 [DAMIT No. 2259] (middle), and DAMIT model 2 [DAMIT No. 2260] (bottom).}
    \label{models1}
\end{figure}

\begin{figure}[ht]
    \centering
    \includegraphics[width=0.45\linewidth]{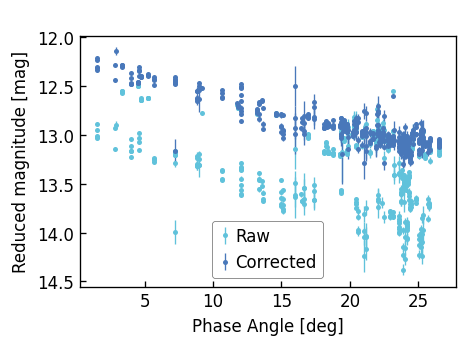}
    \includegraphics[width=0.45\linewidth]{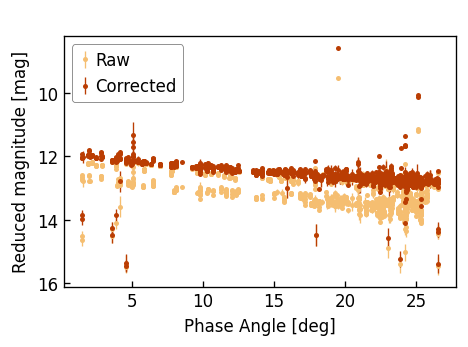}     \\
    \includegraphics[width=0.45\linewidth]{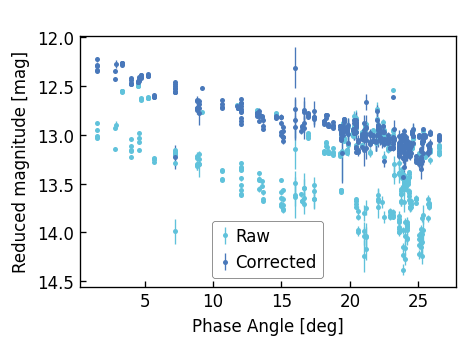}
    \includegraphics[width=0.45\linewidth]{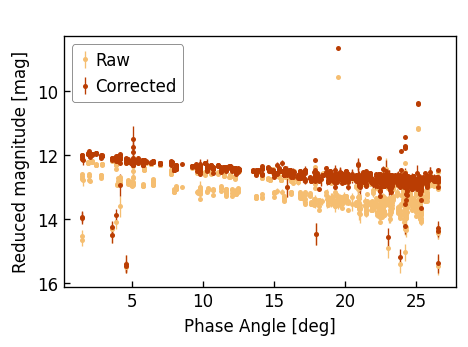}     \\
    \includegraphics[width=0.45\linewidth]{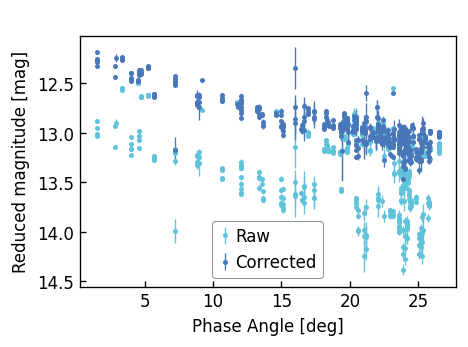}
    \includegraphics[width=0.45\linewidth]{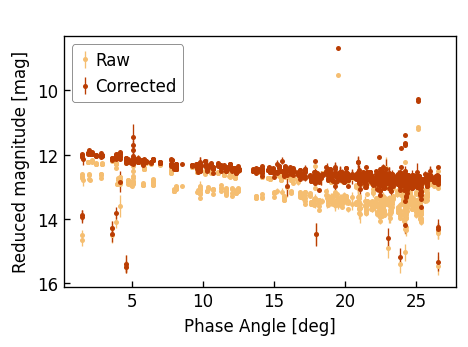}     \\    
    \caption{Data normalization for asteroid (1998) Titius. Black points represent the raw photometric data, while red points show the model- and geometry-corrected photometry reduced to pole-on geometry. The left panels correspond to the ATLAS c filter and the right panels to the o filter. Rows show results for the following shape models: Cellinoid (top), DAMIT model 1 [DAMIT No. 4777] (middle), and DAMIT model 2 [DAMIT No. 4778] (bottom).}
    \label{models2}
\end{figure}

The key objective is to fit the phase function $\Phi(\alpha)$ to photometric data corrected for rotational modulation, shape and apparition effects. To achieve this, we use an empirical approach. Specifically, we employ the SAGE algorithm (Shaping Asteroids with Genetic Evolution; \citep{bartczak2018shaping}) to compute model magnitudes $\delta$ corresponding to the times and observing geometries of the measurements using known spin and shape models—either derived in-house or obtained from public databases such as DAMIT\footnote{\url{https://astro.troja.mff.cuni.cz/projects/damit/}} or ISAM\footnote{\url{http://isam.astro.amu.edu.pl/}}. For the purposes of this work, the number of shape parameters in SAGE method was limited by using a Cellinoid model \citep{cellino1989asteroid}. The shape model parameters were determined based on observational data from the ATLAS sky survey in the c and o filters for phase angles greater than 8 degrees (that is, from observations in the linear regime of phase curve to avoid modelling challenges related to oppostion effect). All calculations were performed using the gaiaathome.eu\footnote{\url{http://gaiaathome.eu/}} service, utilising the computing resources of volunteers. The difference between the model magnitudes and the pole-on-orientation model magnitude is subtracted from the original observed data, thereby normalizing all reduced photometry across apparitions to a common pole-on geometry. Therefore, the fitted absolute magnitudes and phase curve parameters are now directly comparable between the different objects. Figures~\ref{models1} and \ref{models2} show the effect of normalization for the Cellinoid-based model and the models available in the DAMIT database for asteroids (1182) Ilona and (1998) Titius. For Ilona, a substantial reduction in scatter is observed, with the Cellinoid model yielding a significantly greater improvement compared to DAMIT models (pole solutions 1 and 2). For Titius, the reduction in scatter is moderate and largely independent of the chosen model. In both cases, the reduction from a few different apparitions to a common geometry is clearly visible.
Both cases include outliers with large uncertainties; however, these are removed in later processing steps.

The magnitude differences are computed from the given asteroid's known shape, spin state (rotation period $P$ and spin axis orientation), and the observing geometry (unit vectors toward the Sun $\vec{S}$ and observer $\vec{E}$), its disk-integrated brightness can be computed by summing the contributions of all surface facets that are both illuminated and visible. The surface is typically represented as a polyhedral mesh consisting of $N$ triangular facets with known normals $\vec{n}_i$ and areas $A_i$. For each facet, the cosine of the incidence and emission angles are calculated as
\[
\mu_{0,i} = \vec{S} \cdot \vec{n}_i, \quad \mu_i = \vec{E} \cdot \vec{n}_i.
\]
Only facets satisfying $\mu_{0,i} > 0$ and $\mu_i > 0$ contribute to the brightness. Shadowing is accounted for by checking if a facet is occluded by other parts of the shape (using z-buffer method in the non-convex case \citep{akenine2019real}).

The scattered light contribution from facet $i$ is computed using a scattering law $S(\mu_i, \mu_{0,i}, \alpha)$, where $\alpha$ is the solar phase angle. A commonly used model combines the Lommel–Seeliger and Lambert laws with a weighting parameter $c$:
\[
S(\mu_i, \mu_{0,i}) = \mu_i \mu_{0,i} \left( \frac{1}{\mu_i + \mu_{0,i}} + c \right).
\]

The total brightness $L$ of the asteroid (in arbitrary units, normalized by distance factors) is then given by
\[
L = \sum_{i=1}^{N} \cdot S(\mu_i, \mu_{0,i}) \cdot A_i,
\]
summing over all illuminated and visible facets not in shadow. This forward model allows for the generation of synthetic lightcurves for any epoch and geometry and is directly utilized to compute the model magnitude differences between the observing and pole-on geometry.

A few caveats should be noted. The correction for rotational and viewing geometry effects is not based on an analytical formulation, but rather on pre-computed numerical models of shape and spin. These models are themselves derived from observational data, typically via lightcurve inversion, and are subject to uncertainties. Consequently, the corrections inherently depend on model assumptions and uncertainties in the input data.

Nevertheless, this method provides a practical and computationally efficient alternative to more complex approaches that perform simultaneous fitting of spin, shape, and phase function parameters. By relying on pre-defined shape and spin models and applying a separate phase curve fitting step to corrected data, the procedure remains less computationally intensive while still capturing the essential photometric behavior. As such, it represents a complementary strategy to those already established in the literature.

\subsection{Phase curve fitting}
\subsubsection{Fitting the $H$, $G_{1}$, $G_{2}$ phase function}

To fit the $H$, $G_1$, $G_2$ photometric phase function \citep{muinonen2010} to the corrected asteroid magnitudes, we employed a non-linear least-squares optimization procedure. The phase function is given by:

\begin{equation}
V(\alpha) = H - 2.5 \log_{10} \left[ G_1 \Phi_1(\alpha) + G_2 \Phi_2(\alpha) + (1 - G_1 - G_2)\Phi_3(\alpha) \right],
\label{eq2}
\end{equation}

where $V(\alpha)$ is the reduced magnitude at phase angle $\alpha$, $H$ is the absolute magnitude, $G_1$, $G_2$ are the parameters of the phase function and $\Phi_1$, $\Phi_2$, and $\Phi_3$ are empirical phase functions constructed using cubic splines. The fitting was performed using the Levenberg–Marquardt algorithm, as implemented in the \texttt{least\_squares} routine from the \texttt{SciPy} package, using its default convergence settings. These include the function tolerance \texttt{ftol} = $10^{-8}$, parameter tolerance \texttt{xtol} = $10^{-8}$, and gradient tolerance \texttt{gtol} = $10^{-10}$. Additionally, the maximum number of function evaluations was set to the default {\texttt{max\_nfev}} = $100  \times$(number of parameters), which in our case equals to $300$. These default values ensure a balance between computational efficiency and convergence precision. An objective error function was defined as the squared difference between the observed corrected magnitudes and those predicted by the model:

\begin{equation}
\chi^2(H, G_1, G_2) = \sum_{i=1}^{N} \left[ \frac{V_{\text{obs},i} - V_{\text{model}}(\alpha_i; H, G_1, G_2)}{\sigma_i} \right]^2,
\end{equation}

where $V_{\text{obs},i}$ and $\sigma_i$ are the corrected magnitude and its uncertainty for the $i$-th observation, respectively. The initial parameter values were set to $H = \text{mag}_{\text{max}}$, $G_1 = 0.5$, and $G_2 = 0.5$. The algorithm returns the Jacobian matrix of partial derivatives evaluated at the optimal parameter values, from which the covariance matrix $\mathbf{C}$ of the fitted parameters is computed as:

\begin{equation}
\mathbf{C} = (\mathbf{J}^T \mathbf{J})^{-1},
\end{equation}

where $\mathbf{J}$ is the Jacobian. The standard uncertainties of each parameter are then given by the square roots of the diagonal elements of $\mathbf{C}$. From the Jacobian matrix returned by the algorithm, the covariance matrix of the fit was computed, and standard uncertainties for each parameter were derived from its diagonal elements. We note that the covariance-based approach tends to slightly overestimate the uncertainty in $H$. A quick test on a subsample of 100 asteroids shows only a small deviation compared to MCMC-derived errors. Although the covariance method is only slightly faster, even this modest gain becomes relevant when dealing with databases containing tens of thousands of objects.

To test the sensitivity of the solution to the initial conditions, we performed a grid-based experiment using a representative sample of asteroids. Initial guesses were drawn from the sets: $H_{\text{init}} = \{10, 15, 20, 25, 30\}$, $G_{1,\text{init}} = \{0.0, 0.25, 0.5, 0.75, 1.0\}$, and $G_{2,\text{init}} = \{0.0, 0.25, 0.5, 0.75, 1.0\}$, yielding $5 \times 5 \times 5 = 125$ unique starting combinations. The results, %summarized in Figure~\ref{intial_guess}, 
demonstrate that the optimization procedure consistently converges to the same solution, regardless of the initial parameter guess. 
%The narrow distributions of deviations confirm the robustness of the fit.
The differences between the values obtained across all trials and their respective means lie within the range of $[-2 \times 10^{-7}, 2 \times 10^{-7}]$ for the $H$, $G_1$, and $G_2$ parameters, and within $[-4 \times 10^{-8}, 4 \times 10^{-8}]$ for their associated uncertainties. These narrow deviations confirm the robustness and stability of the fitting procedure.

% \begin{figure*}
%     \centering
%     \includegraphics[width=5cm]{Plots/IG_H.png}
%     \includegraphics[width=5cm]{Plots/IG_G1.png}
%     \includegraphics[width=5cm]{Plots/IG_G2.png}
%     \includegraphics[width=5cm]{Plots/IG_std_H.png}
%     \includegraphics[width=5cm]{Plots/IG_std_G1.png}
%     \includegraphics[width=5cm]{Plots/IG_std_G2.png}
%     \caption{Distribution of the deviations of the fitted parameters from their mean values, obtained by varying the initial guesses across 125 combinations. The narrow spread around zero highlights the insensitivity of the final parameter estimates to the choice of initial values. The annotated value indicates the total observed range for each parameter.}
%     \label{intial_guess}
% \end{figure*}

\subsubsection{Fitting the $H,\!G_{12}$ phase function}

The $H,\!G_{12}$ phase function \citep{muinonen2010} is widely used to model the apparent visual magnitude $V$ of asteroids as a function of the phase angle $\alpha$ (eq. \ref{eq2}). In the $H,\!G_{12}$ formalism, the disc-integrated brightness $L_0$ is used ($L_0=10^{-0.4 H}$) and the model is reparametrized using a single parameter $G_{12}$, which is mapped into $(G_1, G_2)$ according to a piecewise definition:

\begin{equation}
(G_1, G_2) = 
\begin{cases}
(0.7527 G_{12} + 0.06164, -0.9612 G_{12} + 0.6270), & \text{if } G_{12} < 0.2 \\
(0.9529 G_{12} + 0.02162, -0.6125 G_{12} + 0.5572), & \text{if } G_{12} \geq 0.2.
\end{cases}
\label{g12}
\end{equation}

\begin{figure}
    \centering
    \includegraphics[width=0.5\linewidth]{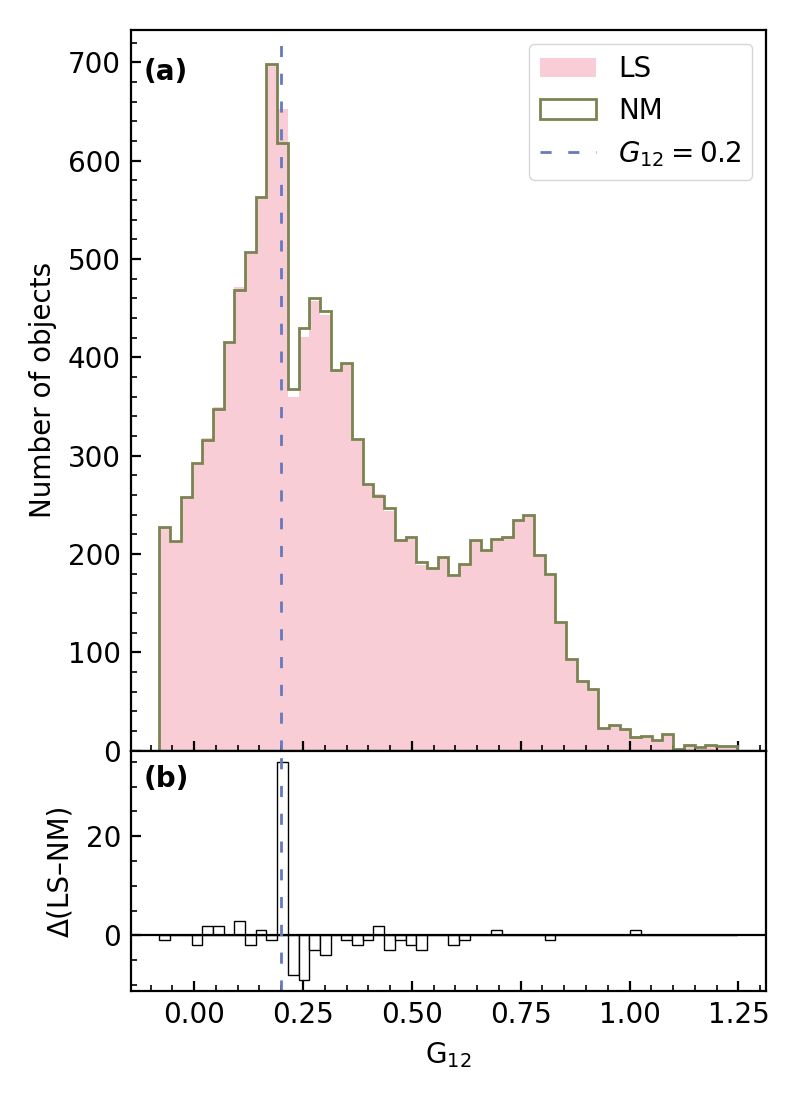}
    \caption{\st{Histogram of fitted $G_{12}$ values using the Nelder–Mead (NM) and least-squares (LS) methods. The dashed vertical line marks $G_{12} = 0.2$, a common convergence point in least-squares optimizations.} Comparison of fitted $G_{12}$ values using the Nelder–Mead (NM) and least-squares (LS) methods. \textbf{(a)} Histograms of the two methods for a sample of 12$\,$114 asteroids; the dashed vertical line marks $G_{12} = 0.2$, a common convergence point in LS optimizations. \textbf{(b)} Difference between the LS and NM distributions (LS $-$ NM), highlighting the small residuals between the methods and confirming that the LS convergence artifact near $G_{12} = 0.2$ is significantly mitigated by the NM method.}
    \label{G12_LS_NM}
\end{figure}

This piecewise definition introduces a discontinuity in the first derivative of the phase function with respect to $G_{12}$ at $G_{12} = 0.2$. While the function $V(\alpha)$ remains continuous, its partial derivative with respect to $G_{12}$ is not:

\begin{equation}
\frac{\partial V}{\partial G_{12}} = -\frac{2.5}{\ln 10} \cdot \frac{\partial f(\alpha, G_1(G_{12}), G_2(G_{12})) / \partial G_{12}}{f(\alpha, G_1, G_2)},
\end{equation}

where the derivative $\partial f / \partial G_{12}$ changes abruptly at $G_{12} = 0.2$ due to the change in $(G_1, G_2)$ dependence. As a result, gradient-based optimization algorithms such as the Levenberg-Marquardt, which underpin standard nonlinear least-squares fitting (e.g., via \texttt{scipy.optimize.least\_squares}), struggle with convergence or may converge to artificial local minima at $G_{12} \approx 0.2$. This phenomenon was empirically observed by \citet{oszkiewicz2011online}, where a significant number of fits converged to the boundary at $G_{12} = 0.2$, resulting in artificial clustering in the fitted parameter distribution. This behavior arises from the optimizer encountering a non-differentiable point in parameter space, which impedes the reliable computation of gradients and leads to unstable convergence. To address this issue, derivative-free optimization algorithms are preferred. In our analysis, we employed the Nelder–Mead simplex method (NM), which relies solely on function evaluations rather than gradients. As a result, it is well-suited to navigate objective functions with discontinuities or non-smooth features. While this method can be slower and potentially less accurate in pinpointing a global minimum compared to gradient-based approaches, it offers more stable and robust performance in fitting the $H$, $G_{12}$ phase function. In Figure \ref{G12_LS_NM}, we show the results of the fitting using the two methods on a sample of 12$\,$114 asteroids. A discrepancy between the two different algorithms at $G_{12} = 0.2$ (dashed line) is evident in the least-square (LS) results, consistent with the convergence issues reported previously of gradient-based optimizers at this boundary. This artifact is significantly mitigated when using the derivative-free NM method. Additionally, the two histograms are nearly identical across all other values, except at $G_{12} = 0.2$, where the NM method shows fewer objects and can resolve values between 0.2 and 0.35, unlike LS, which shows a drop in this range due to many solutions being trapped at 0.2.

To evaluate the sensitivity of the $H,G_{12}$ model parameters to initial conditions, we conducted a systematic grid-based analysis. Initial guesses for the parameters \(L_0\) and \(G_{12}\) were randomly drawn, with \(L_0\) sampled uniformly from the positive range \([0.001, 1.0]\) and \(G_{12}\) from the interval \([0, 1]\). Fits were performed using the two optimization methods under analysis: Nelder--Mead and Levenberg--Marquardt (least squares). The results show that both methods consistently converge to stable solutions, with parameter deviations centered very close to zero. Specifically, for the Nelder--Mead method, the deviations of the \(H\) parameter (computed relative to the mean value as \(H - \langle H \rangle\) ) ranged from [\(-1.76 \times 10^{-9}\), \(2.21 \times 10^{-9}\)], and those of \(G_{12}\) spanned \([-1.22 \times 10^{-8}, 1.27 \times 10^{-8}]\). The corresponding uncertainties varied within similarly narrow bounds, confirming a highly robust fitting procedure. The least-squares method yielded comparable stability, with \(H\) deviations between [\(-9.95 \times 10^{-10}\), \(8.99 \times 10^{-10}\)], and \(G_{12}\) between [\(-7.07 \times 10^{-9}\), \(6.92 \times 10^{-9}\)]. Uncertainties for least-squares fits were also tightly constrained.

\subsubsection{Outlier rejection}

To ensure a robust fit and minimize the influence of outliers, we implemented an outlier rejection procedure based on magnitude uncertainties and residuals relative to a fitted model. This step is critical, as the SAGE correction algorithm does not detect or mitigate observational artifacts that produce anomalous magnitude values. As a result, outliers in the raw data persist after correction and must be removed prior to fitting. A linear-exponential model \citep{kaasalainen2003asteroid} was fitted to the data as a function of phase angle, and residuals were computed. Observations with residuals exceeding 1.5 times the root mean square (RMS) of the residual distribution were then discarded.

\subsubsection{Uncertainty envelopes in HG$_{1}$G$_{2}$ and HG$_{12}$ phase functions}

\begin{figure}
    \centering
    \includegraphics[width=0.5\linewidth]{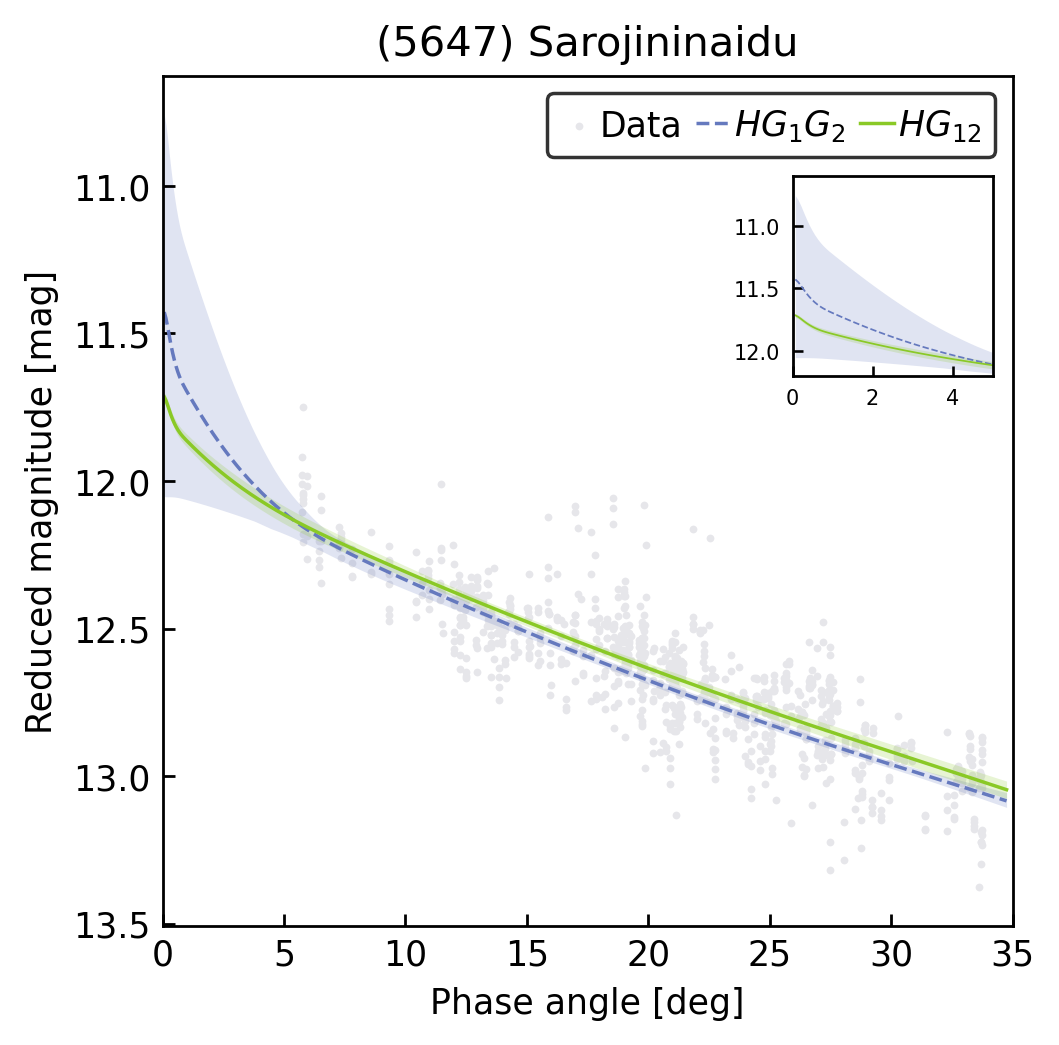}
    \caption{Uncertainty envelopes for the $H,\!G_{1},\!G_{2}$ and $H,\!G_{12}$ phase curves of asteroid (5647) Sarojinianidu.}
    \label{fig:envelops}
\end{figure}

Both the $HG_{1}G_{2}$ and $HG_{12}$ photometric phase functions, introduced by \citet{muinonen2010}, are defined as weighted combinations of three empirical cubic spline basis functions $\Phi_1(\alpha)$, $\Phi_2(\alpha)$, and $\Phi_3(\alpha)$ (eq. \ref{eq2}). The basis functions are normalized such that: $\Phi_1(0) = \Phi_2(0) = \Phi_3(0) = 1,$
which ensures that for all combinations of $G_1$, $G_2$, and $G_3 = 1 - G_1 - G_2$, the model satisfies $V(0) = H$. In the $H,\!G_{12}$ formulation, a single slope parameter $G_{12}$ is mapped linearly to $G_1$ and $G_2$, and $G_3 = 1 - G_1 - G_2$. This results in the same form as the $H,\!G_{1},\!G_{2}$ model, but with a constrained one-dimensional parameter space. Because all basis functions satisfy $\Phi_i(0) = 1$, both models are anchored at opposition: they always return $V(0) = H$ regardless of the slope parameters. However, the behavior of the uncertainty envelope at small phase angles differs significantly between the two models due to their parameterization (see Fig.  \ref{fig:envelops}).

In $H,\!G_{12}$, the mapping from $G_{12}$ to $(G_1, G_2)$ defines a narrow, correlated path in the $(G_1, G_2)$ space (Eq.~\ref{g12}). As a result, the effective variability of the composite phase function near $\alpha = 0^\circ$ is strongly limited, and the derivative of $V(\alpha)$ with respect to $G_{12}$ approaches zero. Consequently, uncertainty envelopes constructed by sampling $G_{12}$ tend to be artificially small near opposition, even in the absence of observational data at small phase angles.

In contrast, the $H,\!G_{1},\!G_{2}$ model allows for independent sampling of $G_1$ and $G_2$ over a broader admissible domain. This greater flexibility permits larger variations in the shape of the phase curve near opposition. Although the constraint $V(0) = H$ still holds, the slope of the phase function at small $\alpha$ may vary substantially across different $(G_1, G_2)$ combinations. As a result, the uncertainty envelope in $HG_{1}G_{2}$ is not artificially constrained near $\alpha = 0^\circ$ and can more accurately reflect the true uncertainty in $H$ when small-angle data are absent.

This distinction is crucial for correctly estimating uncertainties in $H$, especially for datasets with phase angle coverage limited to $\alpha \gtrsim 10^\circ$, large single point uncertainties or data clustered near single phase angle. In such cases, the solution might be degenerate, but the model hides that, because phase curves do not change much under small $G_{12}$ variations. In those cases the $H,\!G_{12}$ model may lead to underestimated errors due to its constrained parameter space. A possible remedy could be to fit the $H,\!G_1,\!G_2$ model, map the $G_1, G_2$ to the $G_{12}$ parameter and use this distribution as a priori for Bayesian analysis of $H,\!G_{12}$ function.

\subsubsection{Improved constrains on the $G_1, G_2$ phase-function parameters}
\label{conHG1G2}

\begin{figure}[ht]
    \centering
    \begin{subfigure}[t]{0.55\linewidth}
        \centering
        \includegraphics[height=5cm]{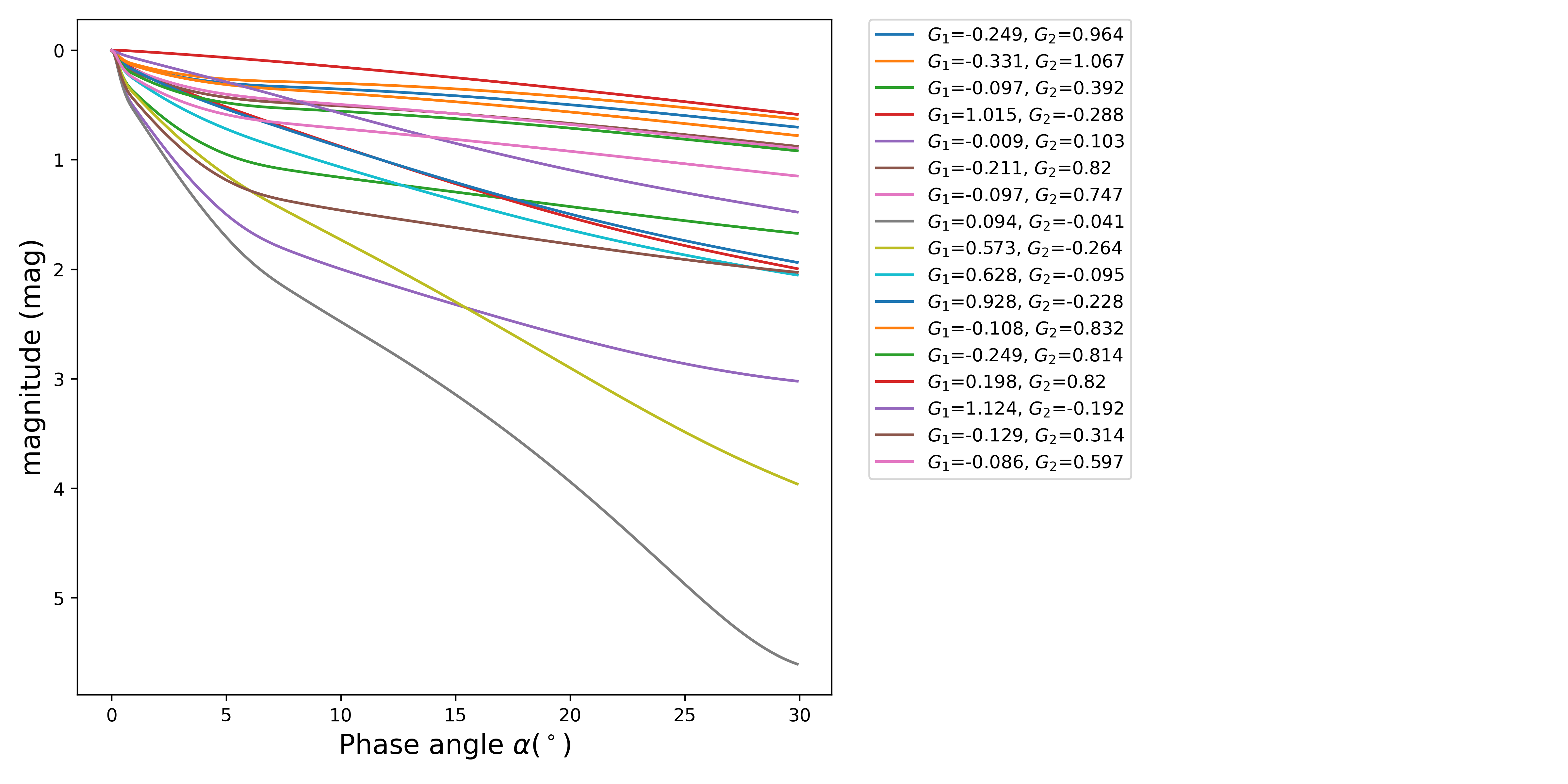}
        \caption{Phase curves with physically valid $G_1$ and $G_2$ parameters that violate the simple conditions defined by \citet{penttila2016h}. The absolute magnitude $H$ is set to zero.}
        \label{outside}
    \end{subfigure}
    \hfill
    \begin{subfigure}[t]{0.4\linewidth}
        \centering
        \includegraphics[height=5cm]{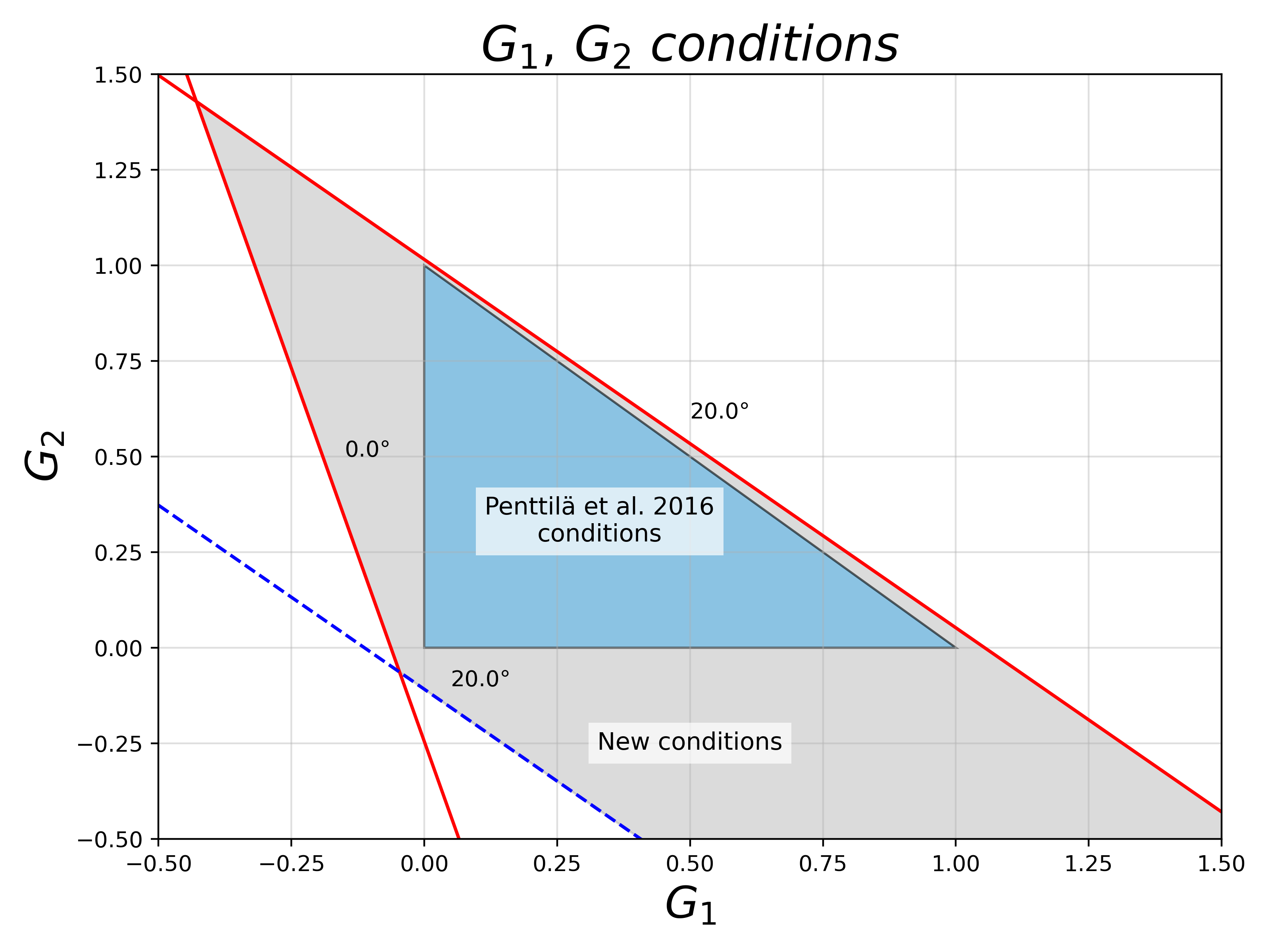}
        \caption{Contour lines of the derivative \( \frac{dV}{d\alpha} = 0 \) at \( 0.001^\circ \) and \( 20^\circ \) (red), and the contour line of \( \frac{dV}{d\alpha} = 1 \) at \( 20^\circ \) (blue).}
        \label{cond}
    \end{subfigure}
    \caption{Left: Phase curves for physical $(G_1, G_2)$ combinations, but violating the \citet{penttila2016h} conditions. Right: Derived constraints visualized in the $(G_1, G_2)$ space.}
    \label{fig:combined_constraints}
\end{figure}

Phase curves in magnitude space are defined by eq. \ref{eq2}. In flux space, the phase function is defined as:

\begin{equation}
    F(\alpha) = a_1 \Phi_1(\alpha) + a_2 \Phi_2(\alpha) + a_3 \Phi_3(\alpha),
\end{equation}

where $a_1$, $a_2$, and $a_3$ are weighting coefficients. To ensure physically meaningful solutions in the case of poor quality data, \citet{penttila2016h} introduced simplified constraints on parameters $a_1$, $a_2$, and $a_3$. By requiring the first derivative of the flux with respect to phase angle to be positive (i.e., to ensure decreasing brightness with increasing phase angle), they proposed that the $a_1$, $a_2$, and $a_3$ coefficients should be restricted to positive values only. This translates into constraints on the parameters $G_1$ and $G_2$: specifically, $G_1$, $G_2 \ge 0$, and $1 - G_1 - G_2 \ge 0$. These conditions ensure that the flux derivative is always positive and hence that the value of magnitude descreases monotonically with phase angle, which corresponds to a physically valid behavior. However, these simplified constraints exclude a subset of valid solutions, in which various combinations of $G_1$, $G_2$ may still produce physically realistic results (see Figure \ref{outside}). To include these cases, we evaluate numerically the first derivative of the magnitude phase curve with respect to phase angle, and require that this derivative is nonnegative: $\frac{dV}{d\alpha} \ge 0 \text{ for all phase angles } \alpha.$ Additionally, to ensure that the derived phase‐curve slopes remain within physically plausible bounds --- and noting that observational studies (e.g., \cite{belskaya2000}) find opposition effect amplitudes of at most $\sim$0.5~mag --- we impose a conservative, yet arbitrary, upper limit on the steepness of the derivative:
$\frac{dV}{d\alpha} \le 1.0\ \mathrm{mag}/\mathrm{deg}\text{ for all phase angles } \alpha$. This choice of 1~mag/{\textdegree} lies well above the empirical maxima, effectively excluding the behaviors of the unphysical steep and unrealistic phase function. The phase function derivative is given by:

\[
\frac{dV}{d\alpha} = - \frac{2.5}{\ln 10} \cdot \frac{G_1 \Phi_1'(\alpha) + G_2 \Phi_2'(\alpha) + \left(1 - G_1 - G_2\right) \Phi_3'(\alpha)}{G_1 \Phi_1(\alpha) + G_2 \Phi_2(\alpha) + \left(1 - G_1 - G_2\right) \Phi_3(\alpha)}.
\]

and is expressed in magnitudes per radian. The derivatives of the spline basis functions take the form $\Phi'(\alpha) = a(\alpha - b)^2 + c(\alpha - d) + e$, where the coefficients $a$, $b$, $c$, and $d$ are listed in Tables~A\ref{coe1}--A\ref{coe3}.

\noindent
We evaluate this derivative numerically across phase-angle intervals between spline nodes, for a grid of $G_1$ and $G_2$ values, both in the range from $-0.5$ to $1.5$. In Figs. A\ref{dVda}, we present a heatmap of the first derivative $\frac{dV}{d\alpha}$ for selected phase angles. We numerically evaluate this derivative across phase angles ranging from $0.001$\textdegree to $150$\textdegree, with $0.001$\textdegree steps, and extract the contour lines corresponding to $\frac{dV}{d\alpha} = 0.0\ \mathrm{mag}/\mathrm{deg}$ and $\frac{dV}{d\alpha} = 1.0\ \mathrm{mag}/\mathrm{deg}$. Among these contours, we identify the most constraining boundaries in the $G_1$–$G_2$ parameter space, which are given by: $G_2 = -3.9038 \cdot G_1 - 0.2445$ (at $\alpha =0.001$\textdegree), $G_2 = -0.9635 \cdot G_1 + 1.0157$ (at $\alpha = 20$\textdegree) and 
$G_2 = -0.9624 \cdot G_1 - 0.1083$  (at $\alpha = 20$\textdegree for $\frac{dV}{dt}=1$). These constraint curves are plotted in Fig. \ref{cond} in comparison to \cite{penttila2016h} conditions. This leads to refined constraints on the parameters \( G_1 \) and \( G_2 \), which can be expressed as:

\begin{equation}
\begin{aligned}
G_1 &\ge -0.429 \\
G_2 &\le 1.429 \\
G_2 &\ge -3.9038 \cdot G_1 - 0.2445 \text{ mandatory condtion deriving from } \frac{dV}{d\alpha} \ge 0 \text{ mag/deg at } 0.001^\circ  \\
G_2 &\le -0.9635 \cdot G_1 + 1.0157 \text{ mandatory condtion deriving from } \frac{dV}{d\alpha} \ge 0 \text{ mag/deg at } 20^\circ \\
G_2 & \ge -0.9624 \cdot G_1 - 0.1083 \text{ (optional condition deriving from } \frac{dV}{d\alpha} \le 1 \text{ mag/deg )}
\label{conditions}
\end{aligned}
\end{equation}

Generally, the permitted parameter space (Fig. \ref{cond}) predominantly includes cases where both $G_1$ and $G_2$ are positive, or where one parameter is positive and the other negative. A small region of the parameter space also allows for both parameters to be negative.

\subsubsection{Improved constrains on the $G_{12}$ phase-curve parameter}
\label{conHG12}

The $H,\!G_{12}$ phase function, proposed by \citet{muinonen2010}, provides a single-parameter empirical model of asteroid brightness as a function of phase angle $\alpha$. It is defined by eqs. \ref{eq2} and \ref{g12}. Using the same constraints, the as for the $H,\!G_1,\!G_2$ phase function, we require that the derivative of the phase function with respect to $\alpha$ remains non-negative. Taking the derivative of the phase function gives:

\begin{equation}
\frac{dV}{d\alpha} = - \frac{2.5}{\ln 10} \cdot \frac{G_1(G_{12}) \Phi_1'(\alpha) + G_2(G_{12}) \Phi_2'(\alpha) + \left(1 - G_1(G_{12}) - G_2(G_{12})\right) \Phi_3'(\alpha)}{G_1(G_{12}) \Phi_1(\alpha) + G_2(G_{12}) \Phi_2(\alpha) + \left(1 - G_1(G_{12}) - G_2(G_{12})\right) \Phi_3(\alpha)}.
\label{dVdalpha-2}
\end{equation}

To identify the domain where this condition is satisfied, we numerically evaluated the derivative on a grid of phase angles and $G_{12}$ values. The results are shown in Fig. A\ref{fig:HG12_derivative}, where the red contour indicates the locus where $dV/d\alpha = 0$. The phase function becomes non-physical for values outside the interval: $-0.08 \leq G_{12} \leq 1.256$ as the derivative turns negative, implying that the modeled brightness increases with phase angle - an unphysical behavior. This interval therefore defines the range of physically admissible values of $G_{12}$ across all phase angles (0\textdegree-150\textdegree). The condition on $\frac{dV}{d\alpha}<1$ mag/deg is not required, as it is already satisfied by all solutions within the specified limits (established numerically - see Fig. \ref{fig:HG12_derivative}).

\subsubsection{Improved constrains on the $G^*_{12}$ phase-curve parameter}
\label{conG12star}
The $G^*_{12}$ parameter was introduced by \cite{penttila2016h} as an improvement over the $H,\!G_{12}$ phase function, after new data revealed that the linear trend between the $G_1$ and $G_2$ parameters continues for S- and M-type asteroids. Thus, for most objects (except D- and E-types), this relationship can be approximated as linear. The parametric relation between $G^*_{12}$ and $G_1$ and $G_2$ is $(G_1, G_2) = (0,\, 0.53513350) + G^*_{12} \times (0.84293649,\, 0.53513350)$. The derivative of the $H, G^*_{12}$ phase function with respect to the phase angle is analogous to eq. \ref{dVdalpha-2}, differing only in that $G_{12}$ is replaced by $G^*_{12}$ along with its corresponding parametric relation. Using the same methodology as for the other two phase functions, we find that the permissible parameter range is $-0.29 \le G^*_{12} \le 1.6979$ (see the corresponding Fig. \ref{G12star}).

\section{Results}\label{res}

\begin{figure}
    \centering
    \includegraphics[width=0.5\linewidth]{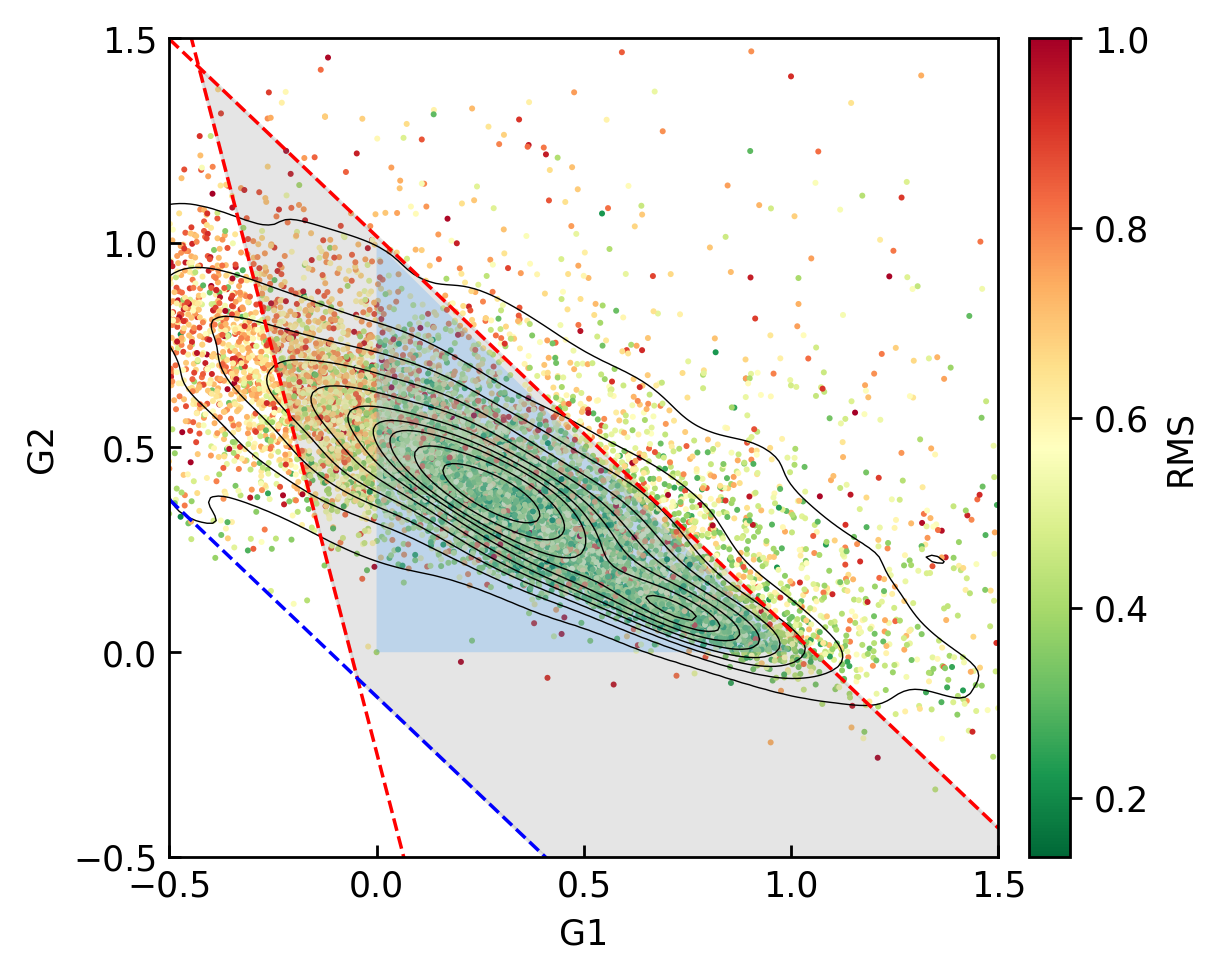}
    \caption{Scatter plot of $G_1$ vs. $G_2$ from the Cellinoid model data, colored by RMS values, with overlaid density contours highlighting regions of higher object concentration. This model is shown as it has the largest sample, with other models producing similar results. Red and blue dashed lines mark boundary conditions based on $\frac{dV}{d\alpha}$ contours at specified phase angles, defining regions of physically plausible parameter values. No objects lie below the blue contour ($\frac{dV}{d\alpha}|{\alpha=20^\circ}=1$), while several exceed the red contours ($\frac{dV}{d\alpha}|{\alpha=20^\circ}=0$) or fall within the enclosed triangular region. Outliers typically correspond to poor-quality photometric data with high scatter (red dots indicating high RMS values).}
    \label{g1g2-scatter}
\end{figure}

We determined reliable phase curve parameters for 9$,$744 out of
15$,$371 objects observed in the cyan filter and 10$,$788 out of
15$,$343 objects in the orange filter. That means, 61\% objects observed in cyan and 70\% objects observed in orange satisfy the post-fit conditions defined in Equation~\ref{conditions} for model 1 (Cellinoid, see Fig. \ref{g1g2-scatter}). For model 2 (DAMIT pole solution 1), parameters were determined for %5$,$798/%
8$,$634 objects in cyan and 8$,$911
%/6$,$392% 
in orange, with 67\% and 71\% of fits, respectively, meeting the criteria of Equation~\ref{conditions}. For model 3 (DAMIT pole solution 2), phase curve parameters were obtained for 4$,$388
%/2$,$952% 
objects in cyan and 4$,$388
%/3$,$243% 
in orange, with 67\% and 73\% of fits, respectively, satisfying the same conditions.

Other fitting algorithms report lower success rates under similar conditions; for instance, the Multi-Apparitions algorithm \citep{colazo2025asteroid}, using the same constraints on $G_1$ and $G_2$ as in this work, achieves a success rate of only 57\% in the orange filter. 

\begin{figure}
    \centering
    \includegraphics[width=5cm]{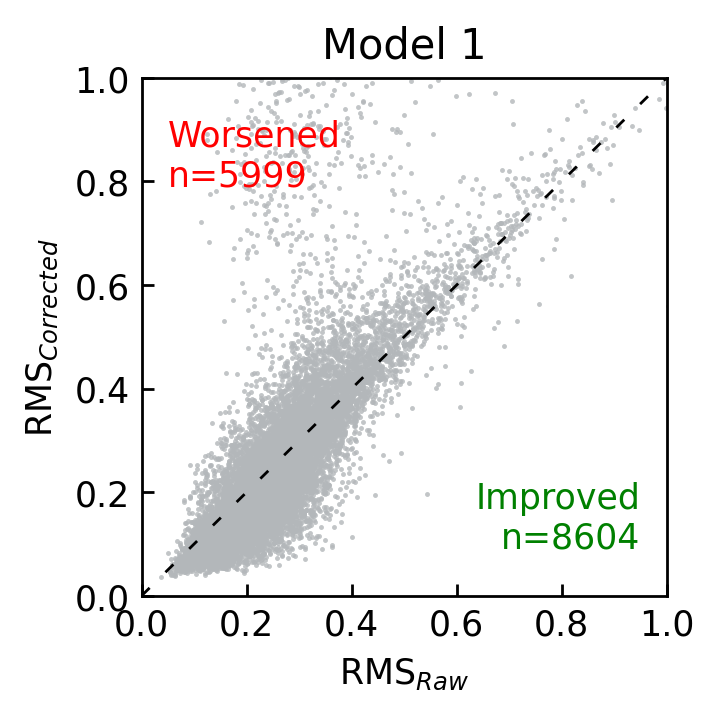}
    \includegraphics[width=5cm]{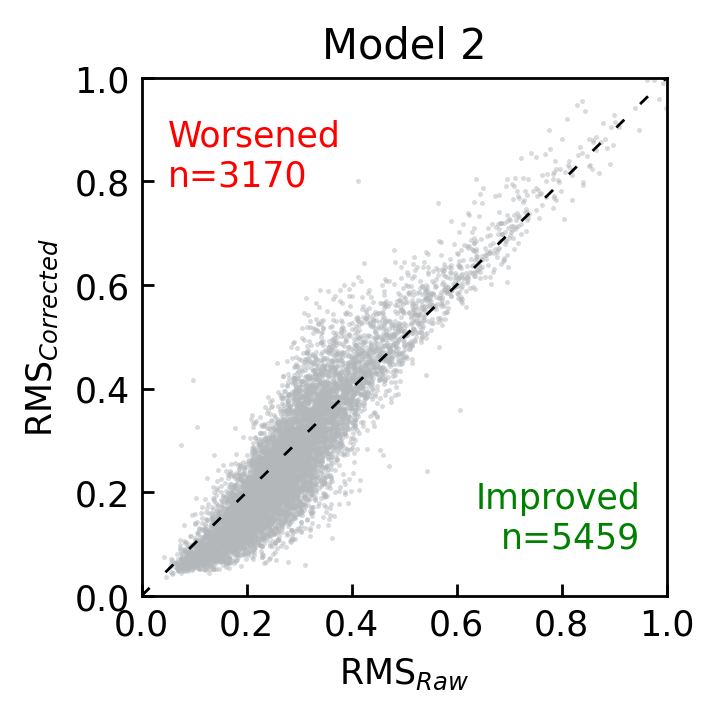}
    \includegraphics[width=5cm]{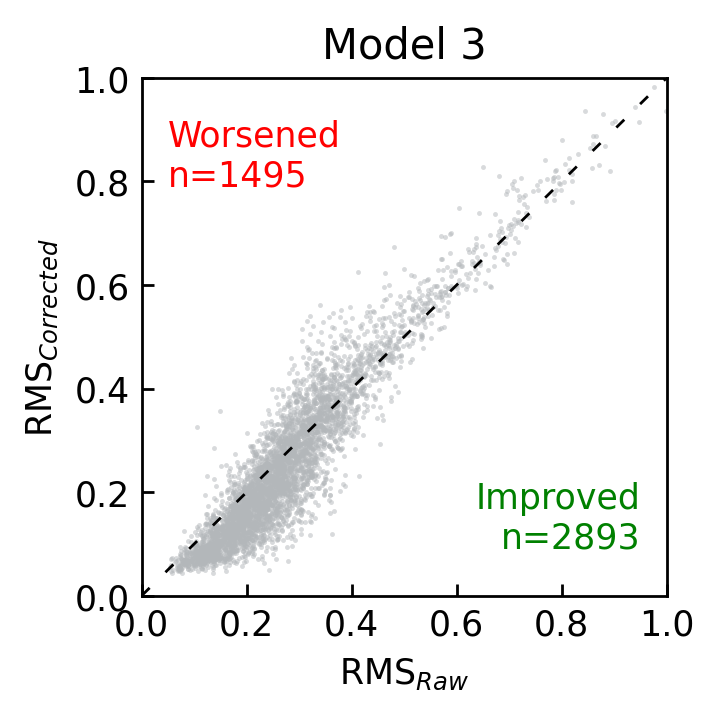}
    \caption{Comparison between RMS values before (RMS$_{Raw}$) and after (RMS$_{Corrected}$) the correction for the three different models discussed in this article. Points below the dashed \(y = x\) line correspond to asteroids whose RMS improved after correction, while points above the line correspond to those whose RMS worsened. Objects lying on the line show no change.}
    \label{rms_comparison}
\end{figure}

In order to analyze the performance of our algorithm, we compared the RMS values before and after the correction for each object by plotting in Figure \ref{rms_comparison} RMS$_{Raw}$ against RMS$_{Corrected}$. The correction improved the RMS for approximately \(54\%\) of the sample for the model 1 (Cellinoid), \(63\%\) for model 2 (DAMIT pole solution 1), and \(65\%\) for the model 3 (DAMIT pole solution 2). We note that, for the Cellinoid data, objects showing improvement typically have more than 1000 observations, suggesting that our algorithm performs better when a large number of data points are available. While the number of observations used for fitting the phase curve is the same across all three models, the accuracy of the underlying shape models—especially for models 2 and 3—likely plays a major role in the fitting performance, since the original data quality and quantity used to derive those shape models are unknown.

To better understand the regions of parameter space where objects fall outside the bounds defined by Equation~\ref{conditions}, we examined the distribution of such cases in $G_1$, $G_2$ space. As shown in Fig.~\ref{g1g2-scatter}, we find almost no objects violating the contion $\frac{dV}{d\alpha}|_{\alpha = 20^\circ} \le 1$ (below the blue dashed line). However, a number of objects violate the $\frac{dV}{d\alpha}|_{\alpha = 20^\circ}, \frac{dV}{d\alpha}|_{\alpha = 0.001^\circ} \ge 0$ conditions (red lines). Upon visual inspection of a subset of these outlier cases, we found that they generally correspond to poor-quality photometric data, often characterized by very large scatter, suggesting that data quality is a key factor in whether fits fall within our defined parameter constraints.

To enable a quantitative model comparison, we compute the Bayesian Information Criterion (BIC) for each fitted phase curve. The BIC provides a model selection criterion that balances goodness-of-fit with model complexity, and is defined as $\mathrm{BIC} = k \cdot \ln(n) - 2 \cdot \ln \mathcal{L}$, where $k$ is the number of free parameters, ($n$) is the number of data points, and ($\mathcal{L}$) is the maximum likelihood of the model. In our case, the fitted model is the $H,\!G_1,\!G_2$ phase function (three parameters). The log-likelihood is computed under the assumption of Gaussian errors, using the standardized residuals between the observed and modeled fluxes. We compute the BIC using the implementation provided by the \texttt{astropy} package. This allows us to directly compare the different shape and spin models available for each asteroid, and objectively identify the preferred one. In Table~\ref{tab:best_model_summary}, we present a summary of preferred models based on BIC values. The model 3 (DAMIT pole solution 2) is the most frequently preferred, being selected in approximately 30\% of the cases. It is followed by the model 1 (Cellinoid), preferred in 22\% of the cases. The model 2 (DAMIT pole solution 1) is selected 11\% of the time. In some cases, two or even all three models yield similar BIC values (within the rounding precision), resulting in no clear preference. 

\begin{table}
\centering
\caption{Summary of preferred models based on BIC values. Model 1 corresponds to the Cellinoid model, model 2 corresponds to the DAMIT pole solution 1 model, and 3 corresponds to the DAMIT pole solution 2 model. The preferred model for each object is identified as the one with the smallest BIC value. In cases where two or more models share the same BIC (within rounding tolerance), the models are listed together, separated by slashes, indicating tied preference.}

\label{tab:best_model_summary}
\begin{tabular}{lrr}
\toprule
Best model & Number of objects & Percentage \\
\midrule
3 & 746 & 30.30\% \\
1 & 547 & 22.22\% \\
1/2& 499 & 20.27\% \\
2 & 289 & 11.74\% \\
1/2/3 & 279 & 11.33\% \\
2/3 & 70 & 2.84\% \\
1/3 & 32 & 1.30\% \\
\bottomrule
\end{tabular}
\end{table}

\begin{figure*}
    \centering
    \includegraphics[width=0.45\linewidth]{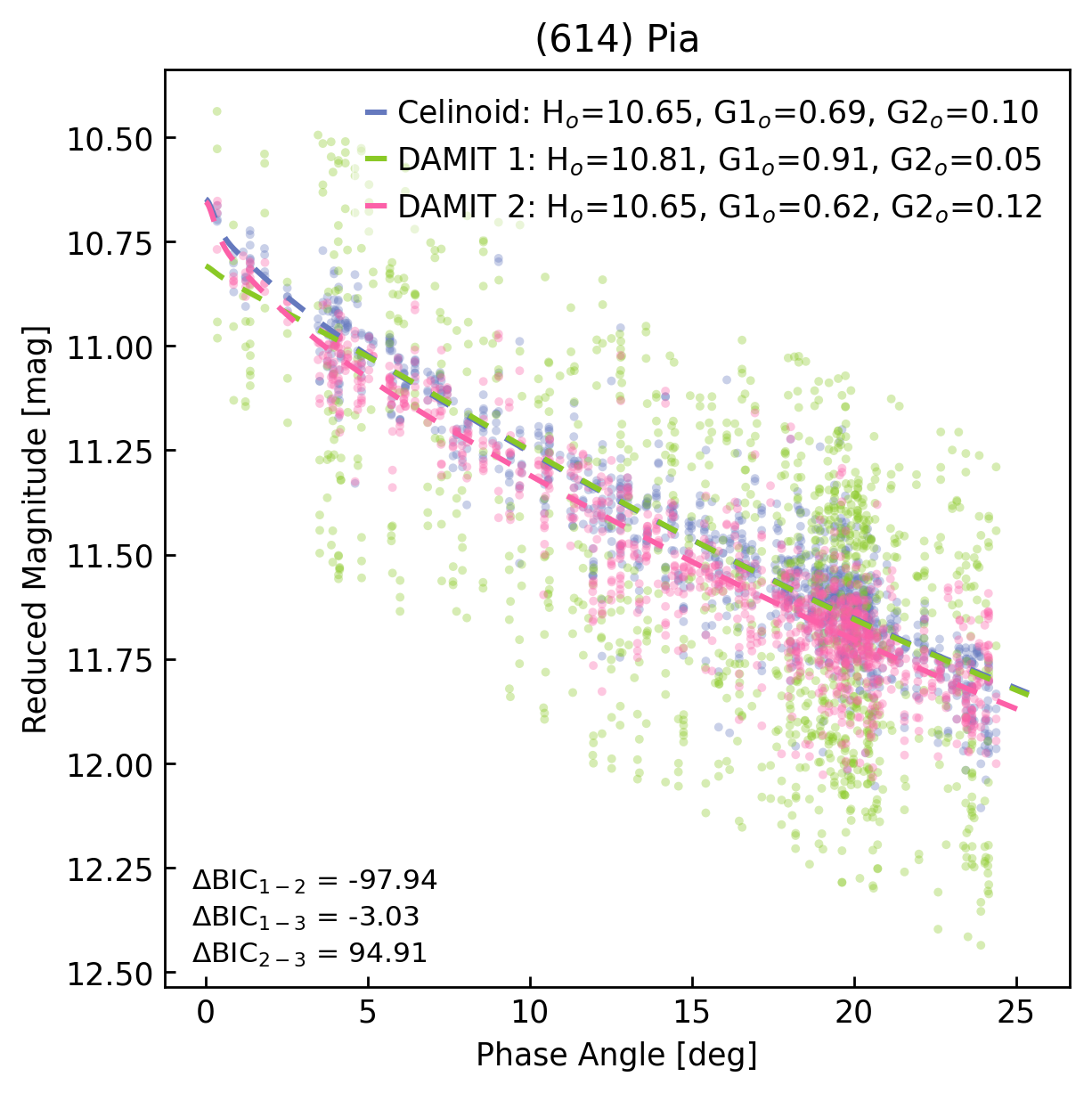}
    \includegraphics[width=0.45\linewidth]{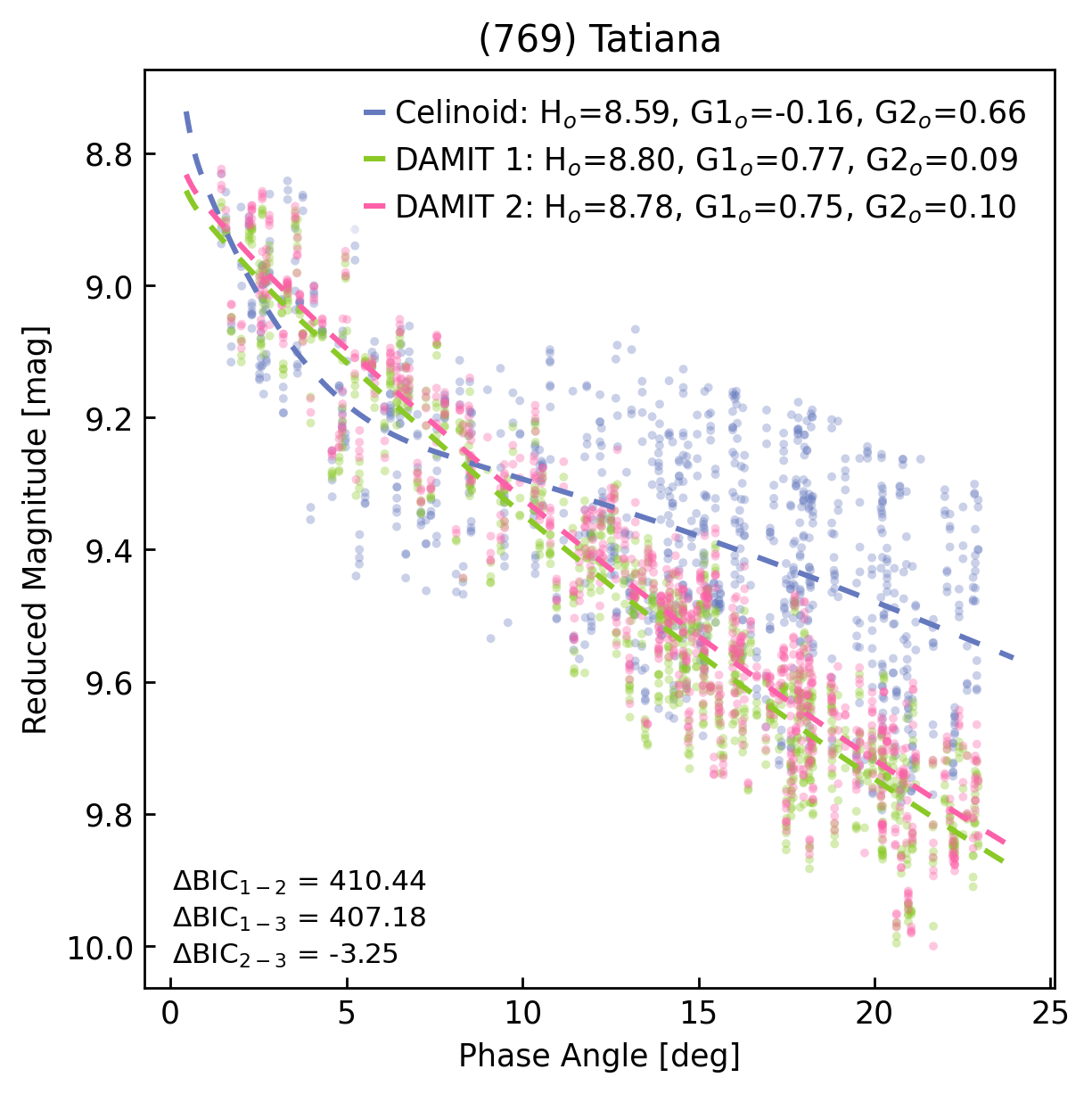}
    \includegraphics[width=0.45\linewidth]{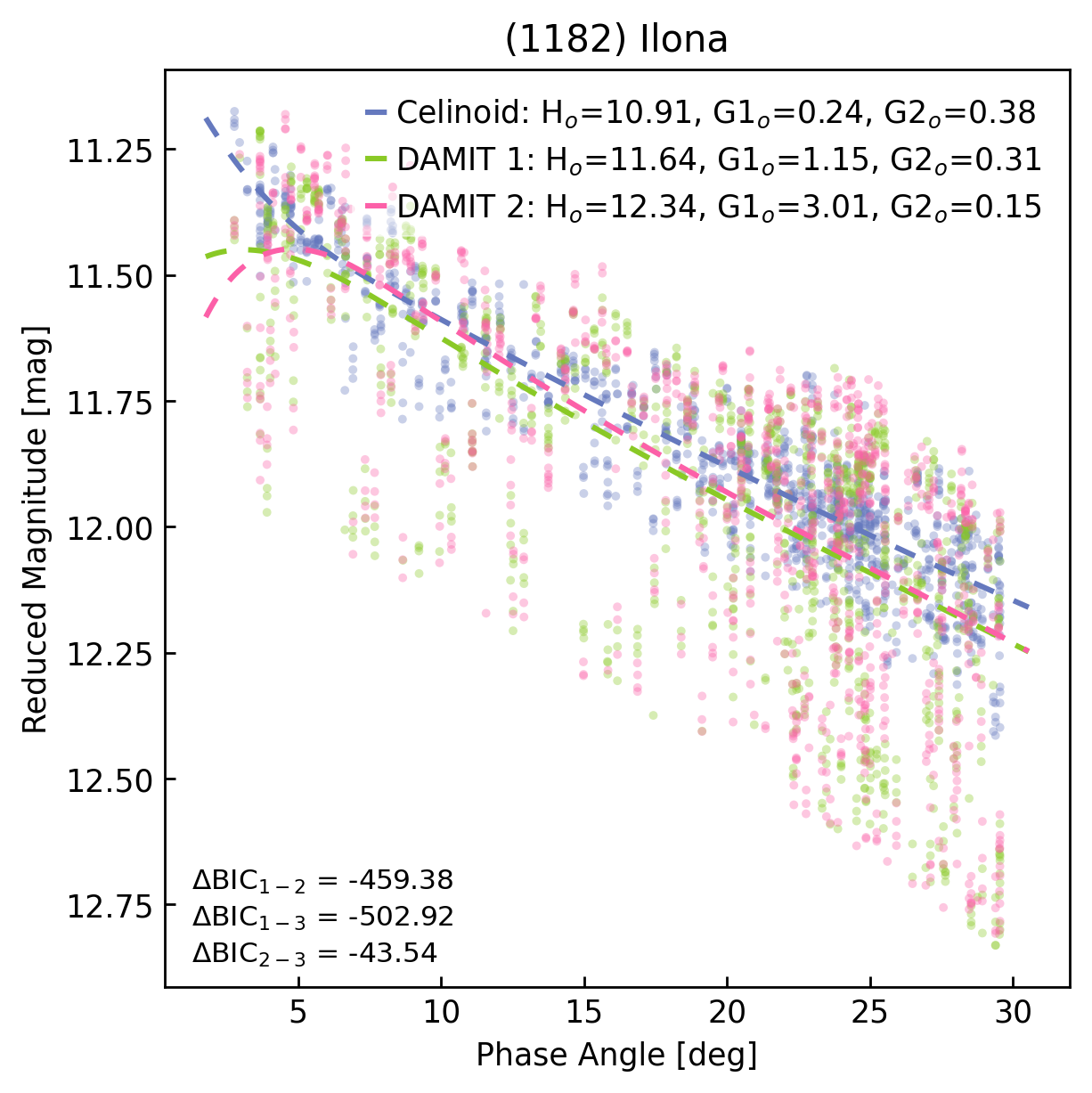}
    \includegraphics[width=0.45\linewidth]{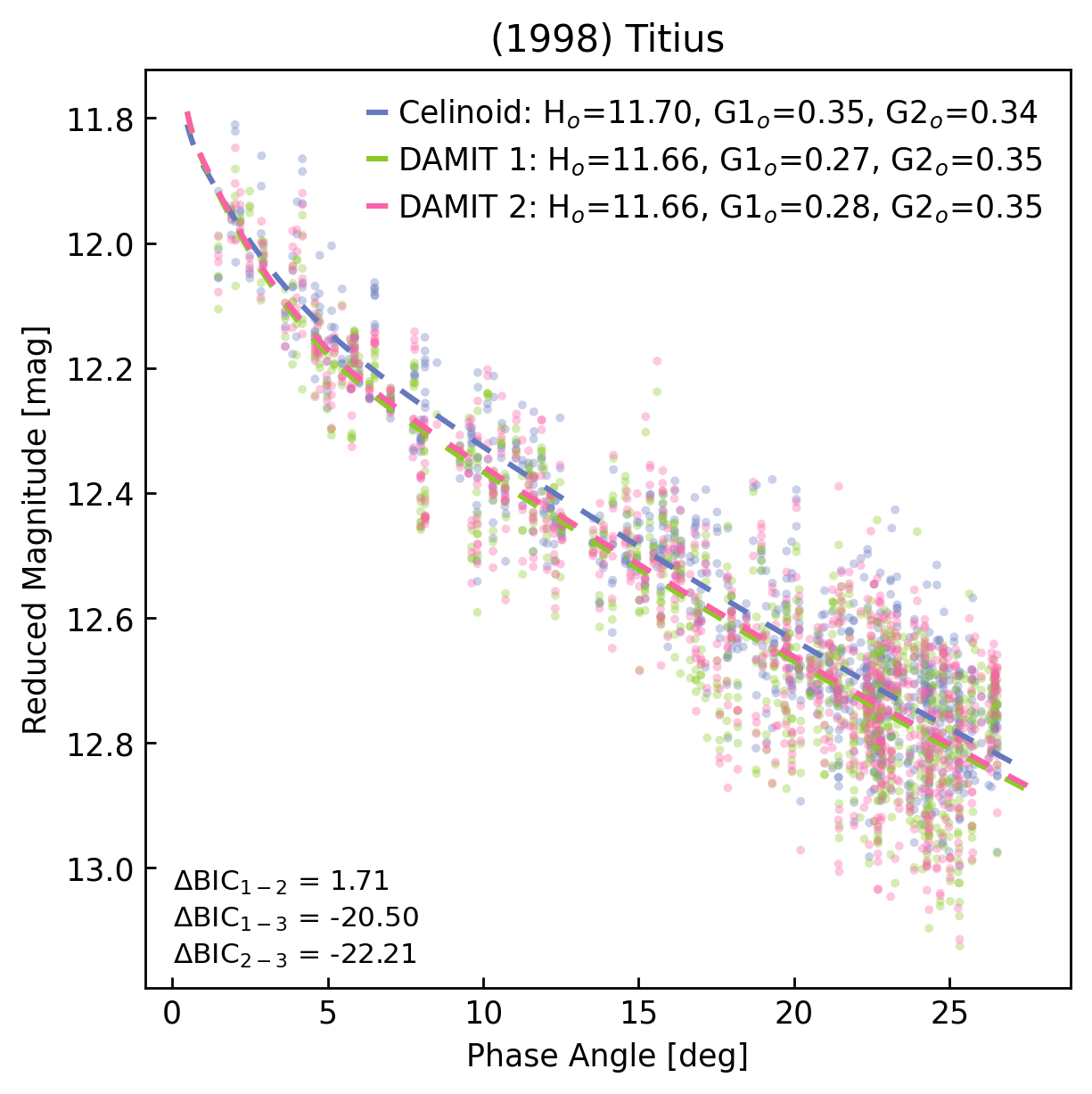}
    \caption{Phase curves for (614) Pia, (769) Tatiana, (1182) Ilona and (1998) Titius. Colors indicate model 1 (Cellinoid), model 2 (DAMIT pole solution 1) and model 3 (DAMIT pole solution 2). The scatter plot shows the corrected magnitudes as a function of phase angle, with the fitted model (dashed line) overlaid. The difference in Bayesian Information Criterion ($\Delta BIC$) is displayed in the corner of the plot, providing a statistical comparison of model performance. For example, $\Delta\mathrm{BIC} = \mathrm{BIC}_1 - \mathrm{BIC}_2 > 0$ indicates that model 2 is statistically favored.}
    \label{phasecurves}
\end{figure*}

As mentioned in Section~\ref{met}, our algorithm takes spin and shape models either computed in-house or from databases such as the ISAM or DAMIT database. For many asteroids, two symmetric pole solutions are available. Fitting phase curves using both models can help identify which pole solution is more plausible, as also explored by \citet{wilawer2024}. In Figure~\ref{phasecurves}, we present four examples of phase curve fits in the ATLAS orange filter: for asteroid (614) Pia, (769) Tatiana, (1182) Ilona, and (1998) Titius. For (614) Pia, the data points produced by model 2 output show significantly more scatter than those from the model 1 and 3 outputs. For (769) Tatiana, the phase curve fit based on the model 1 output displays an unrealistic opposition effect, whereas the fits from both DAMIT model outputs provide a more physically consistent solution. For (1182) Ilona, the fit from the model 1 output reproduces the phase curve more accurately, while those from DAMIT models outputs exhibit an unrealistic opposition effect. These unphysical fits are removed after fitting by applying the conditions described in Sections \ref{conHG1G2}, \ref{conHG12}, \ref{conG12star}. For (1998) Titius, the fits from all three model outputs describe the phase curve well, which is also reflected in the BIC values, all close to zero.

\section{Conclusions}
\label{concl}

We have introduced an empirical method for correcting asteroid phase curves for rotational, shape and geometrical effects using pre-computed spin and shape models. By normalizing sparse photometric data to a pole-on geometry, the approach enables consistent phase-curve fitting across multiple apparitions and improves comparability between objects.

Our work provides:
\begin{itemize}
    \item A robust normalization framework applicable to large survey datasets, demonstrated here with ATLAS photometry for over 15,000 asteroids.
    \item Refined physical constraints on phase-function parameters $G_1$, $G_2$ and $G_{12}$, derived numerically from monotonicity and slope-limit conditions, which extend the permitted solution space beyond earlier formulations while excluding unphysical behaviors.
    \item Algorithmic guidance for stable fitting, including the use of derivative-free methods for $H, G_{12}$ to avoid convergence artifacts at $G_{12} \approx 0.2$.
    \item A model-selection framework using BIC to identify the most plausible spin--shape solution for each asteroid.
\end{itemize}

The method improves RMS residuals for a majority of targets and allows statistical discrimination between competing shape models. While the corrections depend on the accuracy of the adopted spin and shape solutions, they offer a computationally efficient alternative to full spin--shape--phase inversion.

%\section*{Data Availability}

\section*{Author contributions}
DO and PB derived the concept of empirical corrections, and PB provided normalized magnitudes and is responsible for the data normalization section. DO derived the boundary conditions for the phase-curve parameters. MC performed phase curve fitting. DO and MC wrote majority the text, with input from other co-authors.

\section*{Acknowledgments}
DO, PB and MC  were supported by grant No. 2022/45/B/ST9/00267 from the National Science Centre, Poland. PB acknowledges funding through the Spanish Government
retraining plan 'María Zambrano 2021-2023' at the University of Alicante
(ZAMBRANO22-04) and ESA contract No. 4000120180 in the framework of the Permanently Open Call for Proposals under the Polish Industry Incentive Scheme (AO8433). The authors also thank the volunteers of the BOINC Gaia@home project (\url{http://gaiaathome.eu}) for providing computing power used in this research. ChatGPT and Overleaf Writefull were used for language editing and improving the figures. After using those tools, the authors reviewed and edited the content as needed and take full responsibility for the content of the publication. 
\noindent

\bibliographystyle{model5-names.bst}
\bibliography{biblio}

\onecolumn
\appendix
\section{Anex}

\subsection{Coeficients for the derivatives of basis functions}
\FloatBarrier
\vspace{0.5em}
%\begin{table}[Hb!]
\begin{center}
 %   \centering
    \begin{tabular}{|c|c|c|c|c|c|} \hline
    $\alpha$ range (\textdegree) & a & b & c & d & e\\ \hline
    7.5\textdegree - 30\textdegree & -6.812748 & 0.1309 & 6.126412 & 0.1309 & -1.909859 \\
    30\textdegree - 60\textdegree &  -0.348572 & 0.5236 & 0.775692 & 0.5236 & -0.554634 \\
    60\textdegree - 90\textdegree & -0.240592 & 1.0472 & 0.410668 & 1.0472 & -0.244046 \\
    90\textdegree - 120\textdegree & -0.034786 & 1.5708 & 0.158721 & 1.5708 & -0.094980 \\
    120\textdegree -150\textdegree & -0.488589 & 2.0944 & 0.122292 & 2.0944 & -0.021411 \\
    \hline
    \end{tabular}
    \captionsetup{type=table}
    \captionof{table}{Coeficients for $\frac{d\Phi_1}{d\alpha}$.}
    \label{coe1}
\end{center}
%\end{table}

%\begin{table}[Hb!]
\begin{center}
  %  \centering
    \begin{tabular}{|c|c|c|c|c|c|} \hline
    $\alpha$ range (\textdegree) & a & b & c & d & e\\ \hline
    7.5\textdegree - 30\textdegree & 3.274255 & 0.1309 & -1.780058 & 0.1309 & -0.572958 \\
    30\textdegree - 60\textdegree & -0.379534 & 0.5236 & 0.791536 & 0.5236 & -0.767054 \\
    60\textdegree - 90\textdegree & -0.110903 & 1.0472 & 0.394089 & 1.0472 & -0.456658 \\ 
    90\textdegree - 120\textdegree & 0.085536 & 1.5708 & 0.277952 & 1.5708 & -0.280718 \\
    120\textdegree -150\textdegree & -0.294370 & 2.0944 & 0.367525 & 2.0944 & -0.111733\\
    \hline
    \end{tabular}
    \captionsetup{type=table}
    \captionof{table}{Coeficients for $\frac{d\Phi_2}{d\alpha}$.}
    \label{coe2}
\end{center}
%\end{table}

%\begin{table}[Hb!]
 %   \centering
 \begin{center}
    \begin{tabular}{|c|c|c|c|c|c|} \hline
    $\alpha$ range (\textdegree) & a & b & c & d & e\\ \hline
    0\textdegree - 0.3 \textdegree & 2428451.278843 & 0.0000 & -20559.923775 & 0.0000 & -0.106301 \\
    0.3\textdegree - 1.0\textdegree & -192238.573083 & 0.0052 & 4870.758549 & 0.0052 & -41.180439 \\
    1.0\textdegree - 2.0\textdegree & -785.953504 & 0.0175 & 173.484079 & 0.0175 & -10.366915 \\
    2.0\textdegree - 4.0\textdegree & -997.734942 & 0.0349 & 146.049126 & 0.0349 & -7.578461 \\
    4.0\textdegree - 8.0\textdegree & -497.196388 & 0.0698 & 76.394087 & 0.0698 & -3.696095 \\
    8.0\textdegree - 12.0\textdegree & -34.054440 & 0.1396 & 6.972375 & 0.1396 & -0.786057 \\
    12.0\textdegree - 20.0\textdegree & -2.510511 & 0.2094 & 2.217478 & 0.2094 & -0.465270 \\
    20.0\textdegree - 30.0\textdegree & -1.971925 & 0.3491 & 1.516411 & 0.3491 & -0.204595 \\
    \hline
    \end{tabular}
    \captionsetup{type=table}
    \captionof{table}{Coeficients for $\frac{d\Phi_3}{d\alpha}$.}
    \label{coe3}
    \end{center}
%\end{table}

\vspace{9cm}
\FloatBarrier
\subsection{Heatmaps of $\frac{dV}{d\alpha}$ for the $H,G_1,G_2$ phase function}

\begin{figure}[htbp]
\centering

% First row
\begin{subfigure}[b]{0.29\textwidth}
\includegraphics[width=\linewidth]{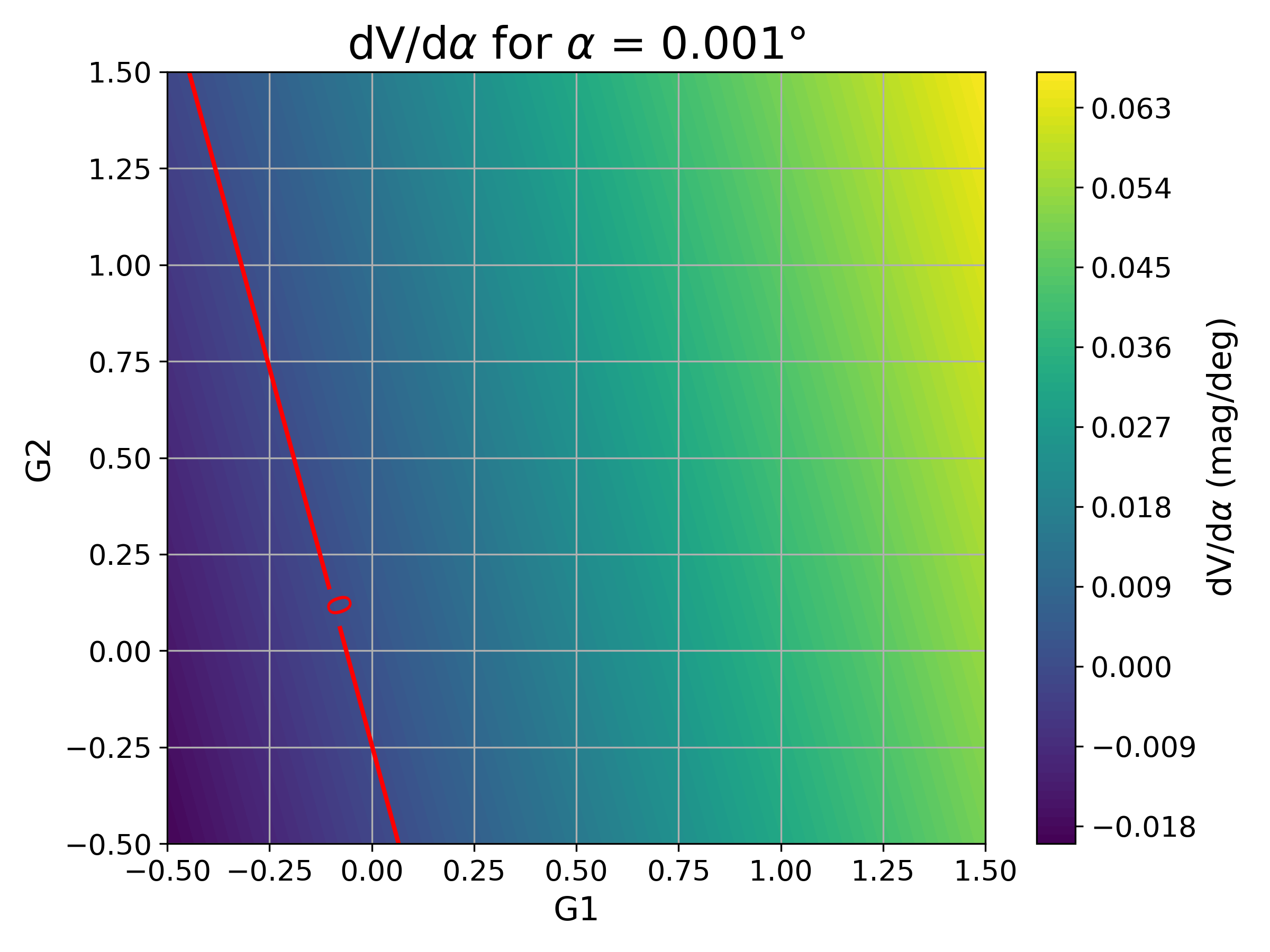}
\end{subfigure}
\hfill
\begin{subfigure}[b]{0.29\textwidth}
\includegraphics[width=\linewidth]{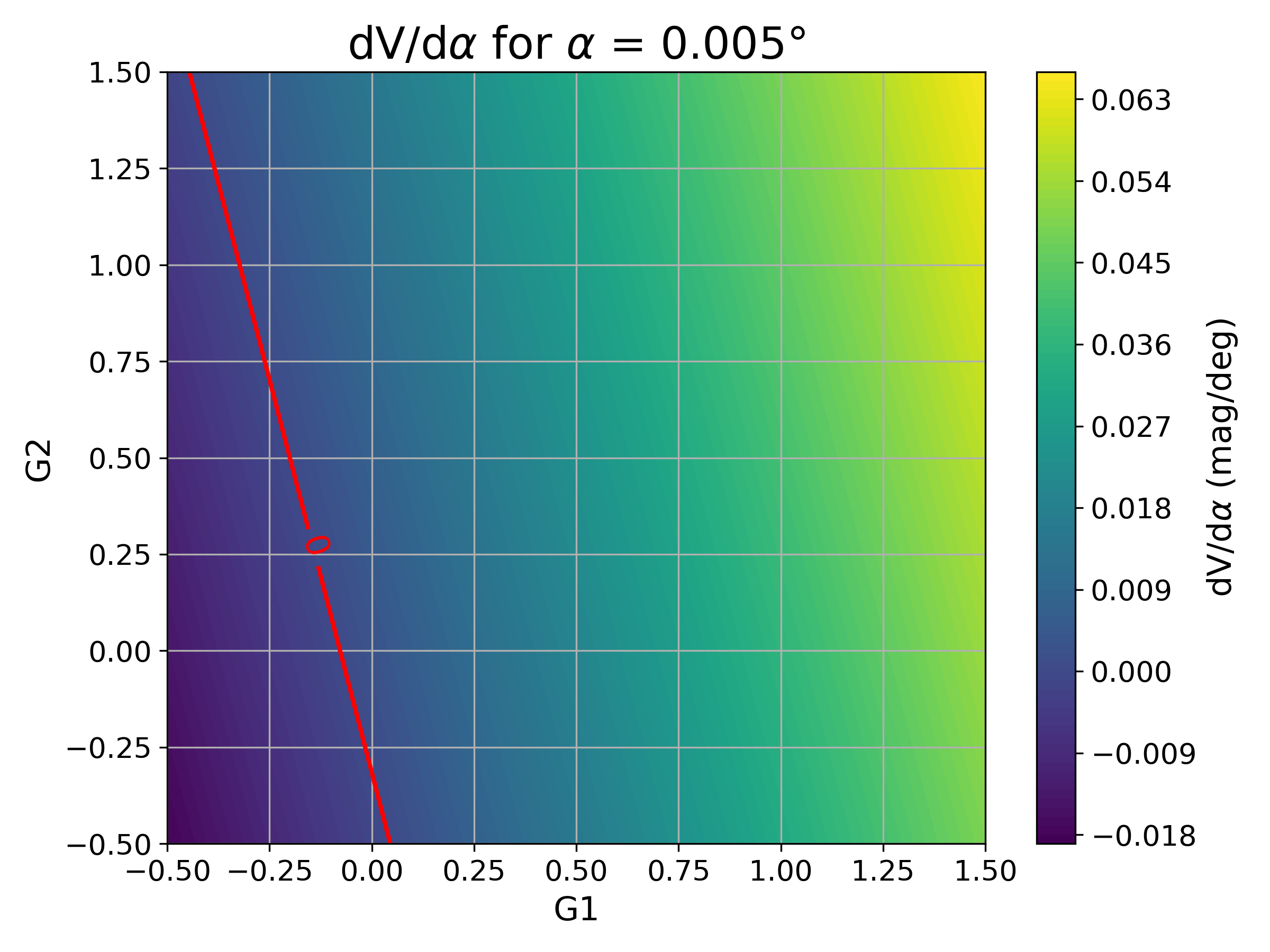}
\end{subfigure}
\hfill
\begin{subfigure}[b]{0.29\textwidth}
\includegraphics[width=\linewidth]{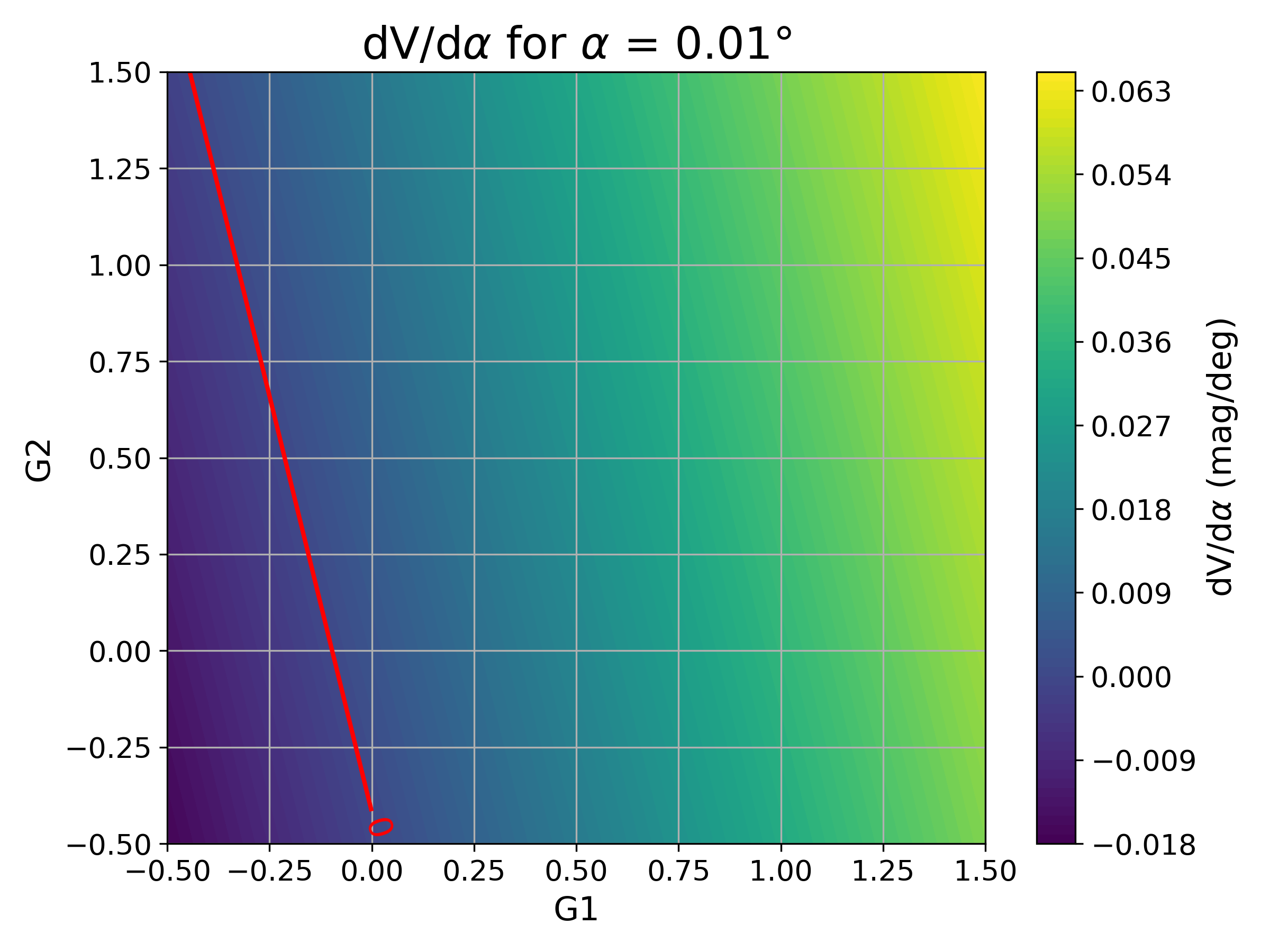}
\end{subfigure}

%\vspace{0.05cm}

% Second row
\begin{subfigure}[b]{0.29\textwidth}
\includegraphics[width=\linewidth]{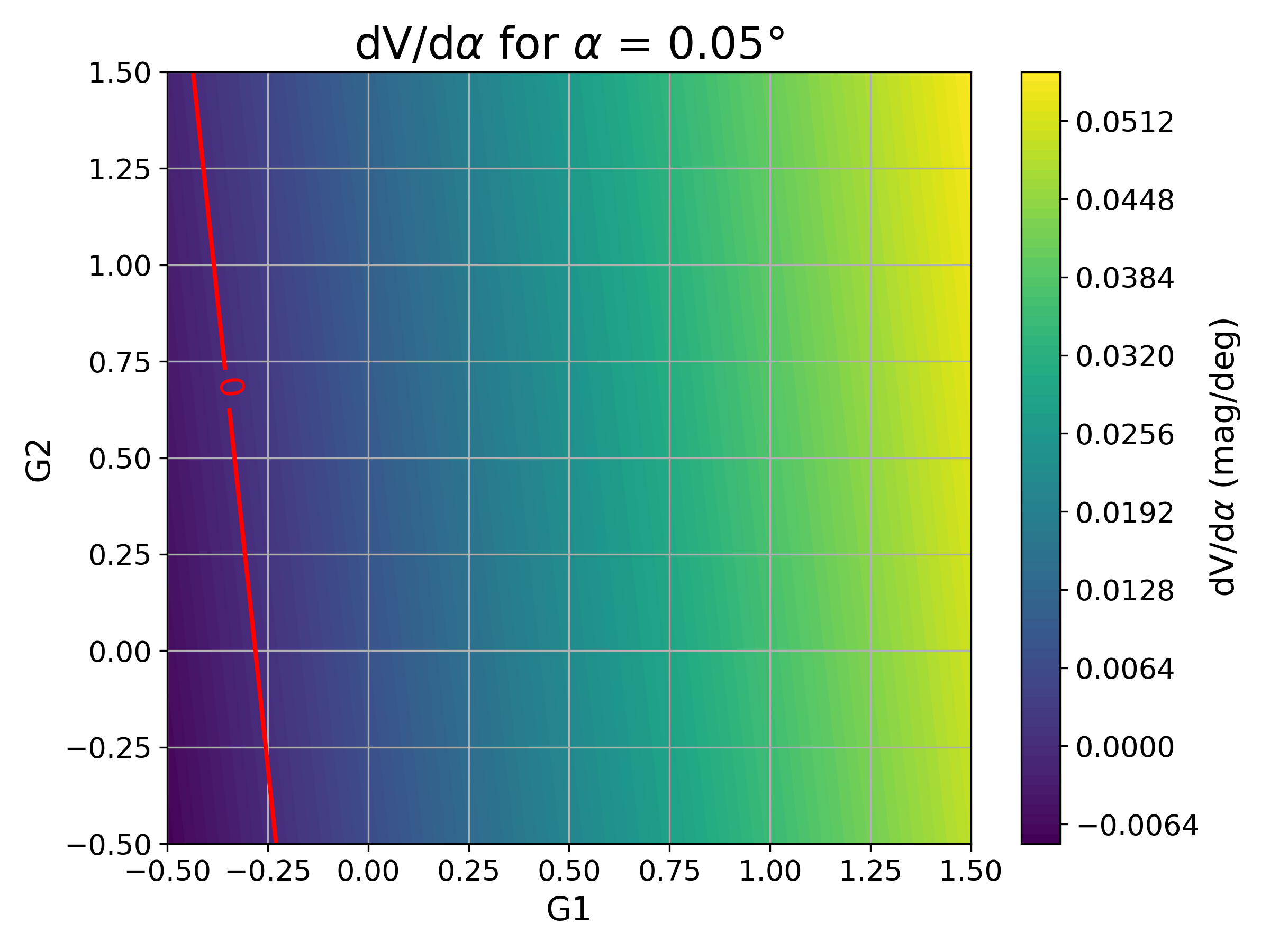}
\end{subfigure}
\hfill
\begin{subfigure}[b]{0.29\textwidth}
\includegraphics[width=\linewidth]{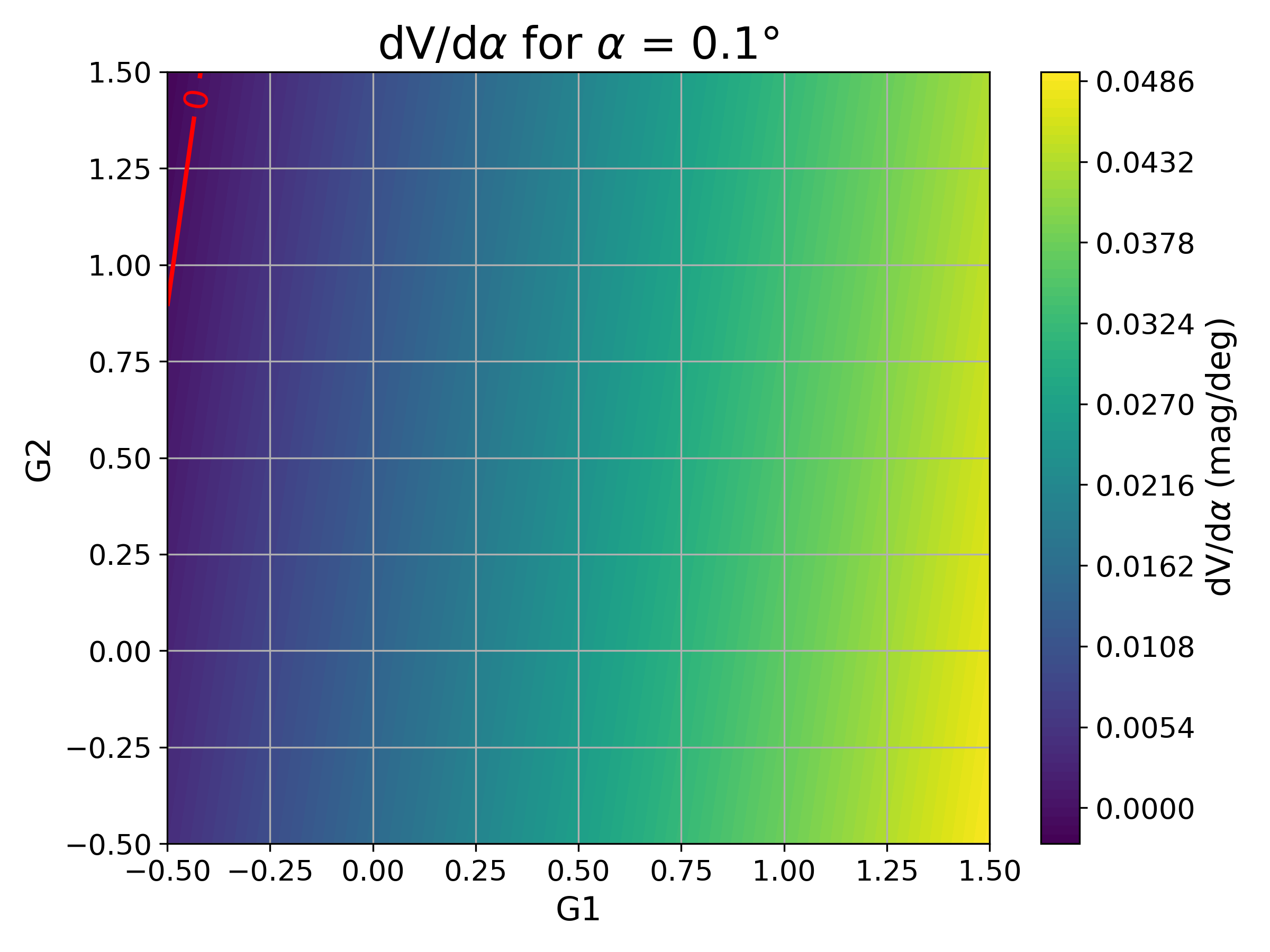}
\end{subfigure}
\hfill
\begin{subfigure}[b]{0.29\textwidth}
\includegraphics[width=\linewidth]{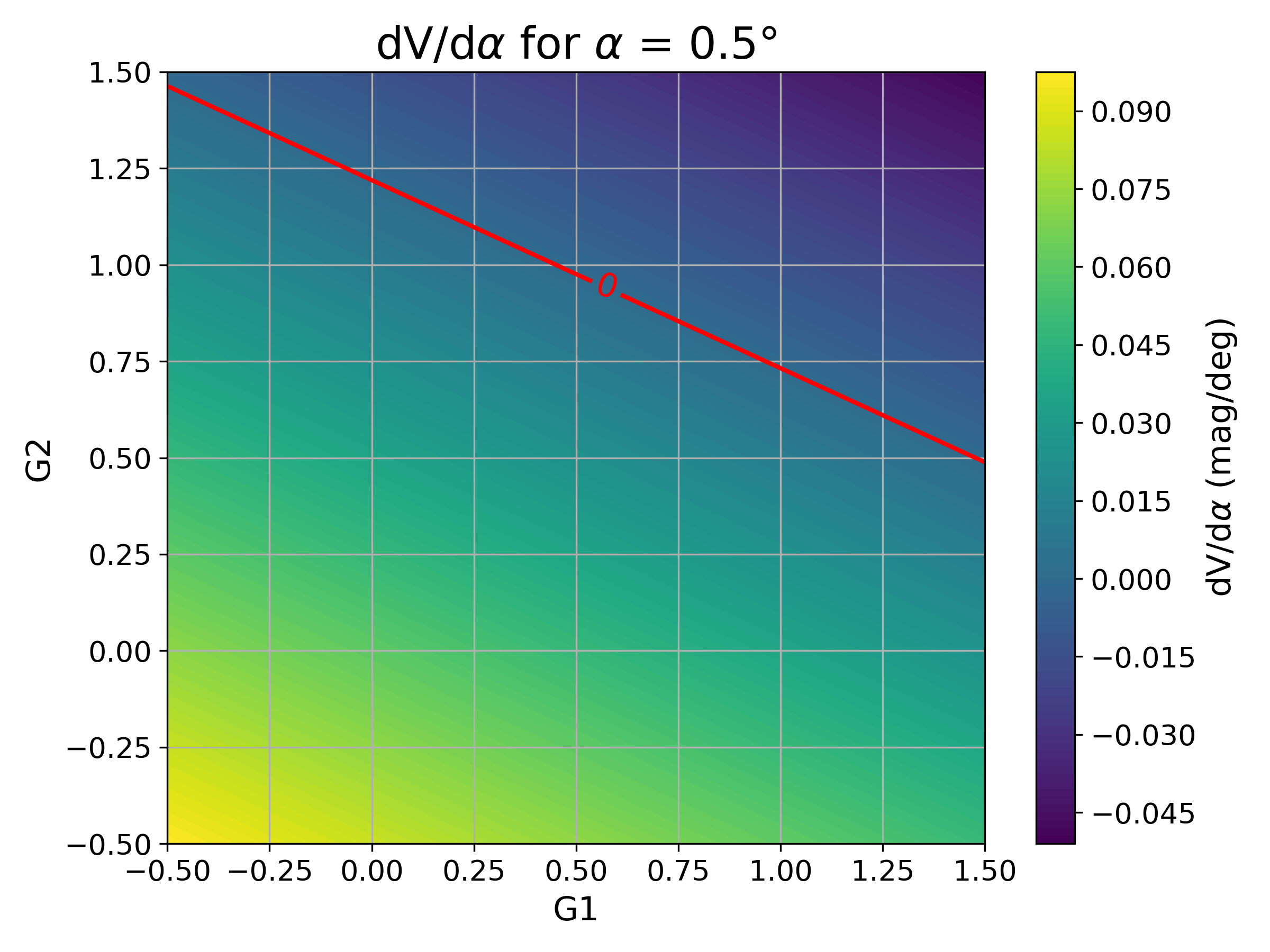}
\end{subfigure}

%\vspace{0.05cm}

% Third row
\begin{subfigure}[b]{0.29\textwidth}
\includegraphics[width=\linewidth]{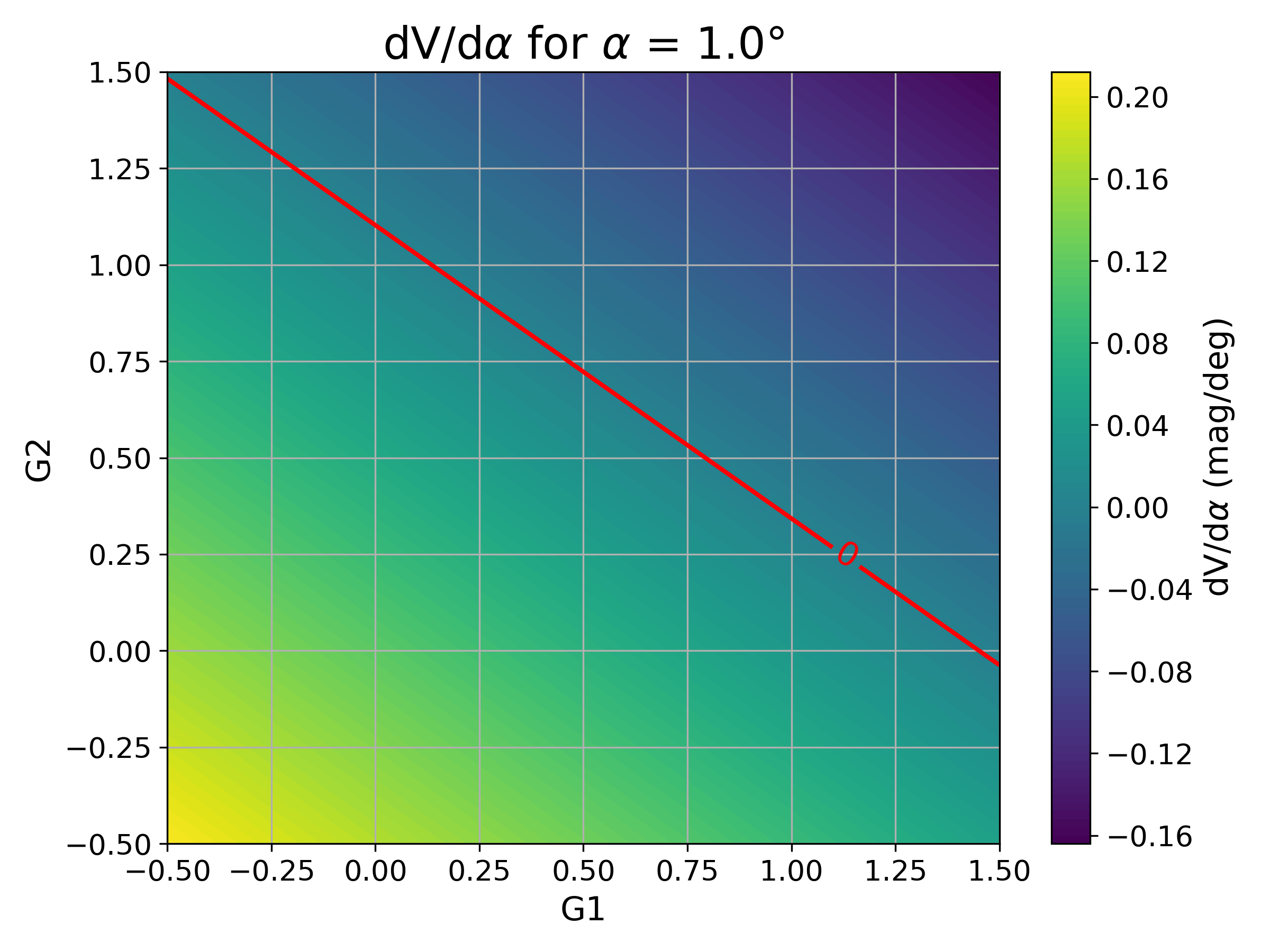}
\end{subfigure}
\hfill
\begin{subfigure}[b]{0.29\textwidth}
\includegraphics[width=\linewidth]{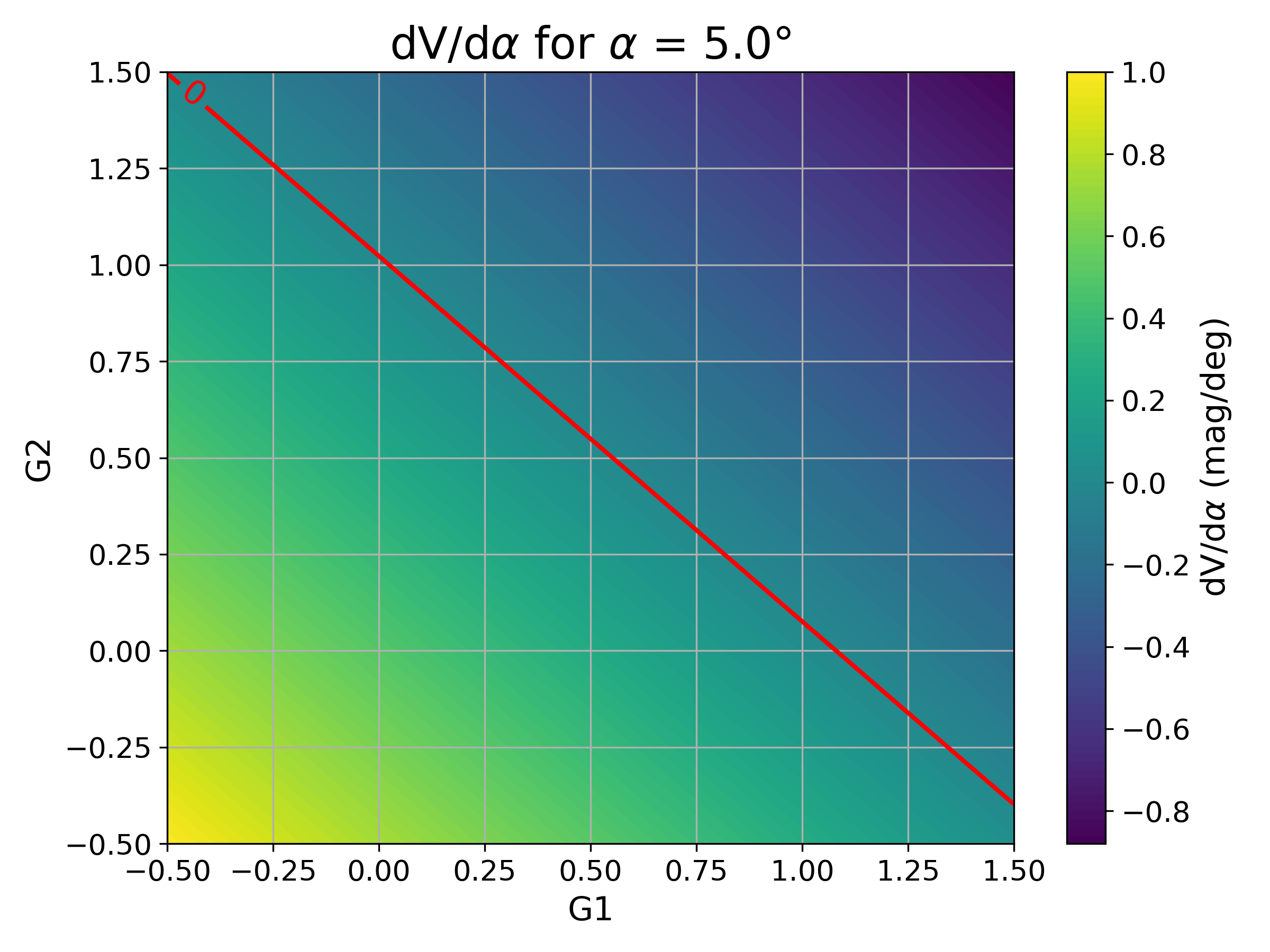}
\end{subfigure}
\hfill
\begin{subfigure}[b]{0.29\textwidth}
\includegraphics[width=\linewidth]{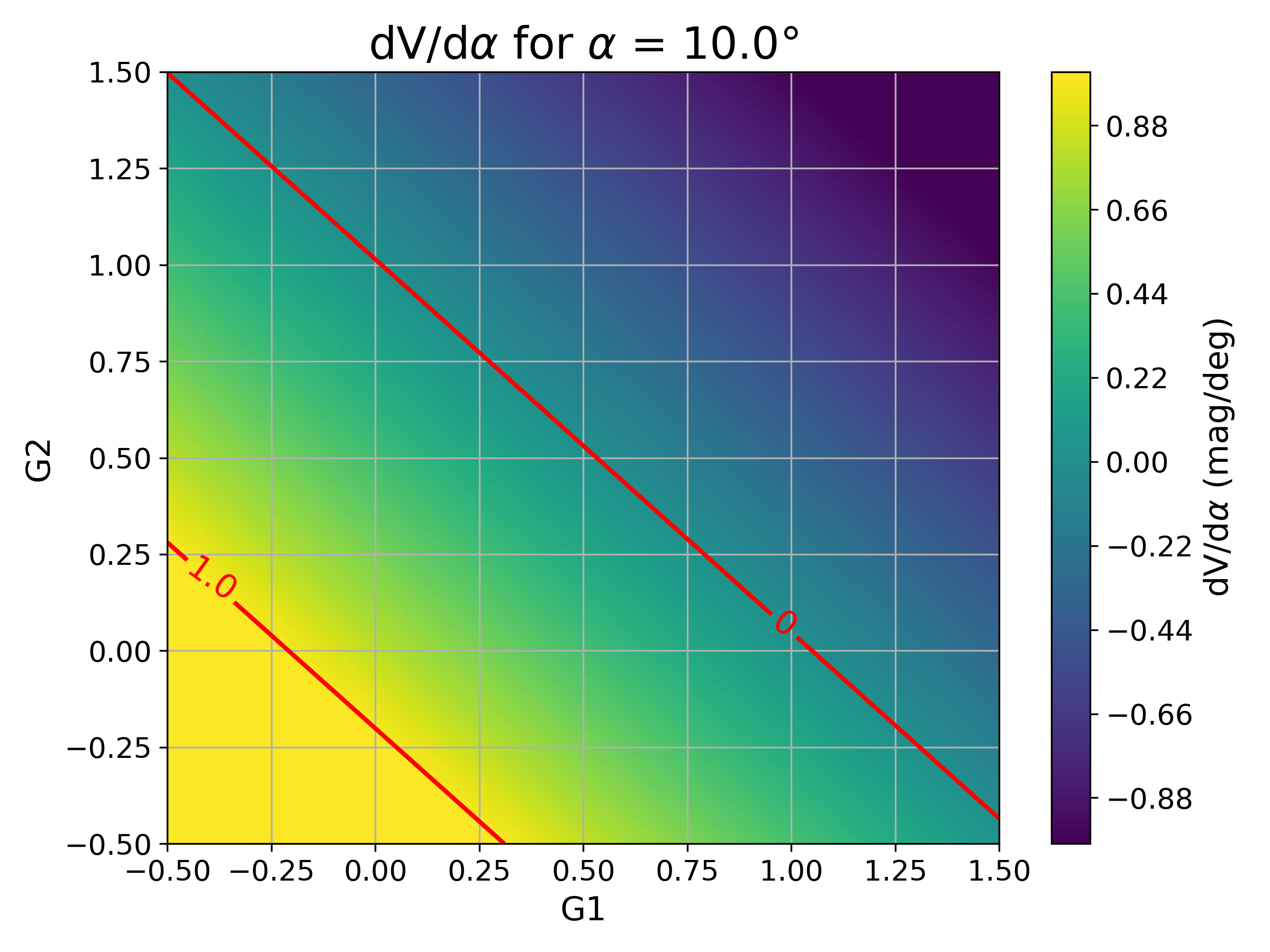}
\end{subfigure}

%\vspace{0.05cm}

% Fourth row
\begin{subfigure}[b]{0.29\textwidth}
\includegraphics[width=\linewidth]{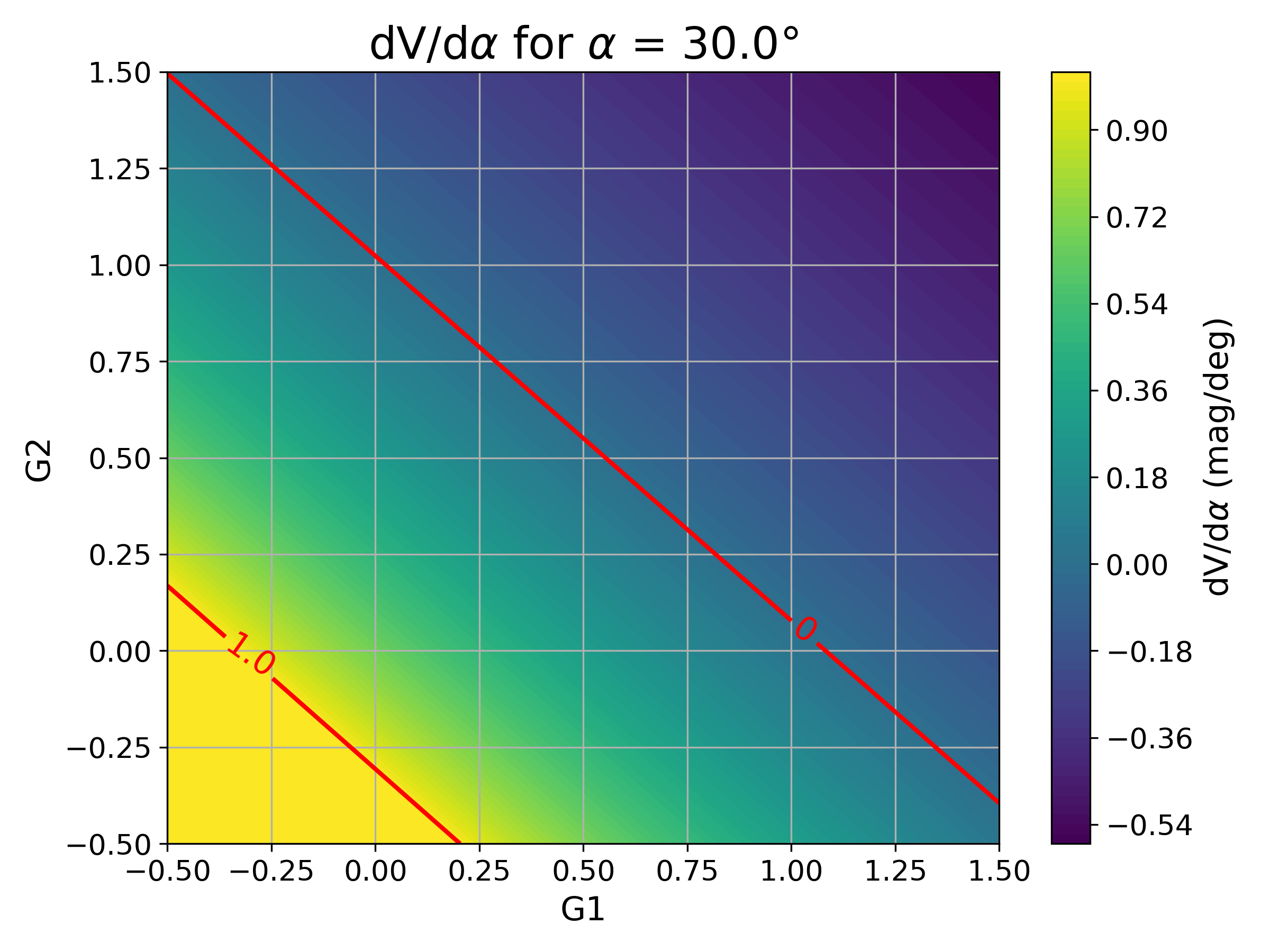}
\end{subfigure}
\hfill
\begin{subfigure}[b]{0.29\textwidth}
\includegraphics[width=\linewidth]{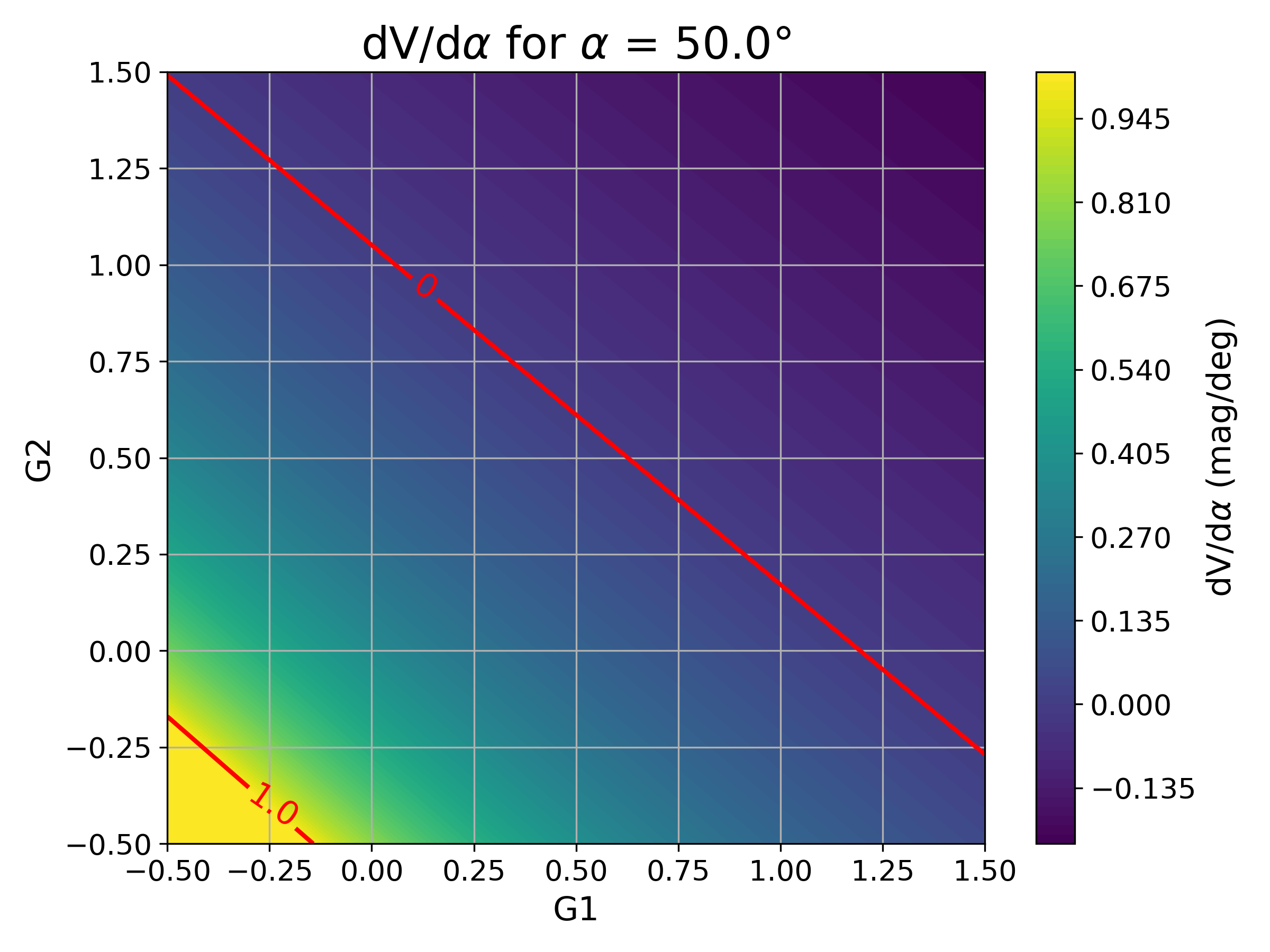}
\end{subfigure}
\hfill
\begin{subfigure}[b]{0.29\textwidth}
\includegraphics[width=\linewidth]{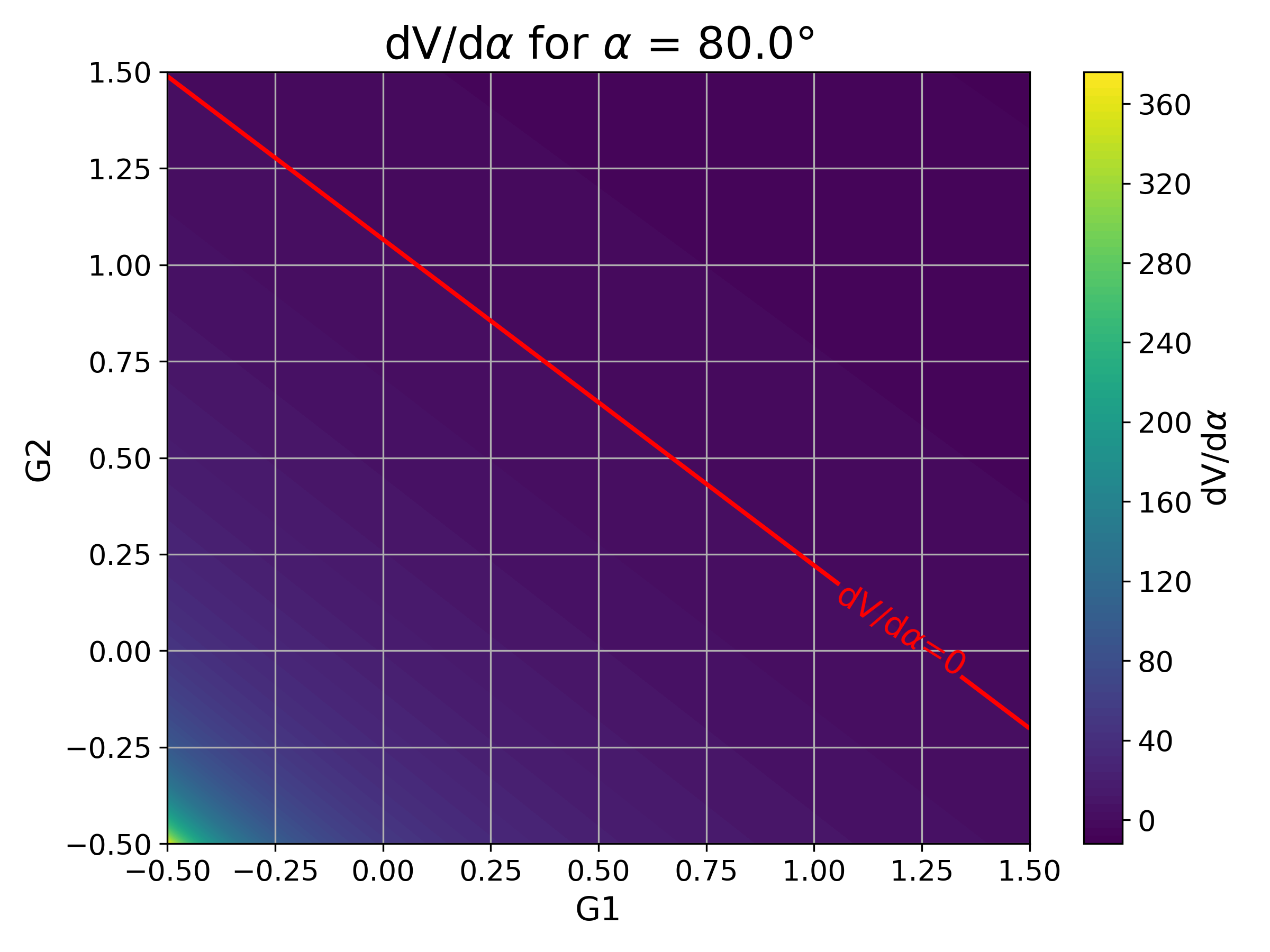}
\end{subfigure}

%\vspace{0.05cm}

% Fifth row
\begin{subfigure}[b]{0.29\textwidth}
\includegraphics[width=\linewidth]{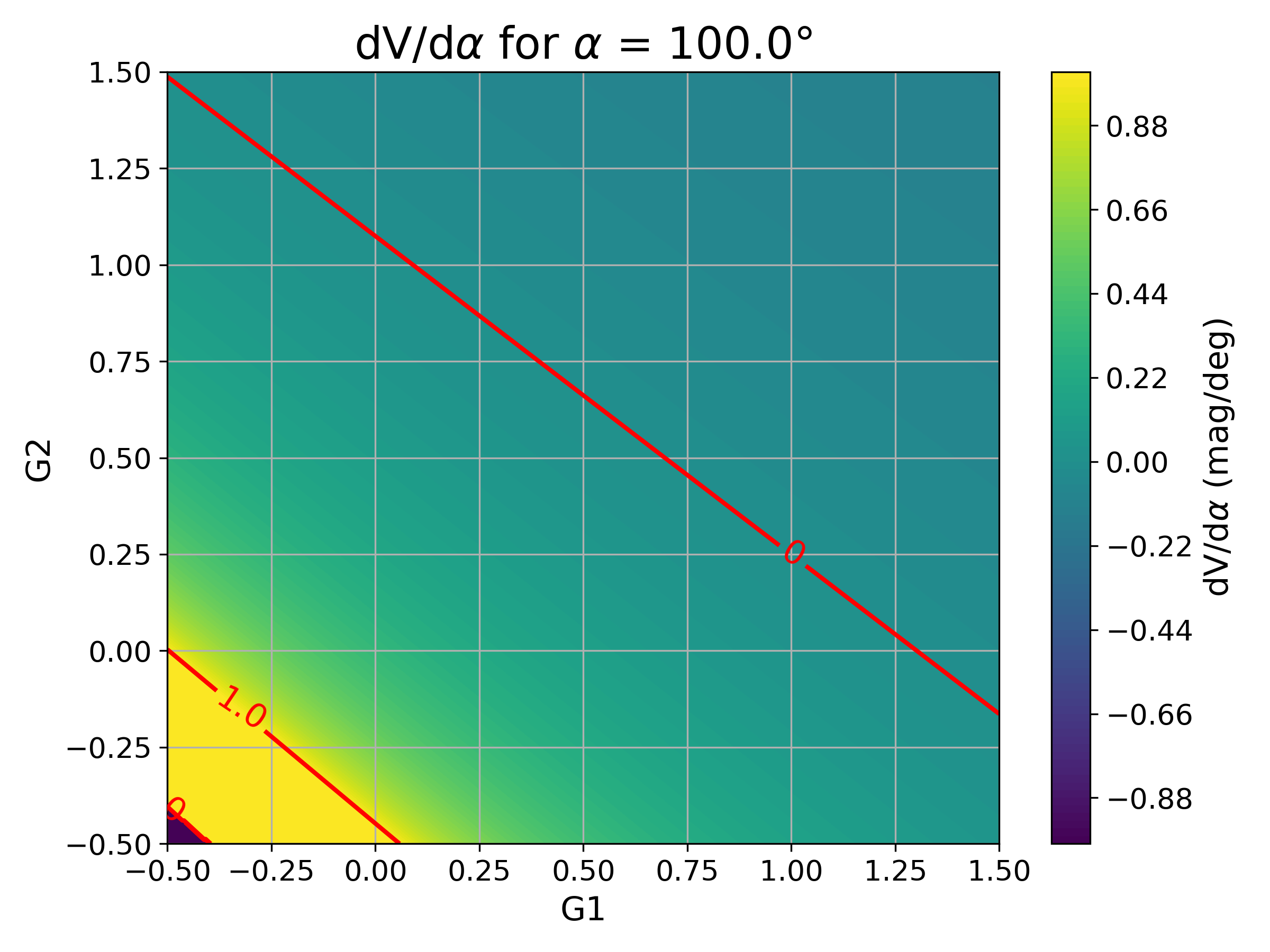}
\end{subfigure}
\hfill
\begin{subfigure}[b]{0.29\textwidth}
\includegraphics[width=\linewidth]{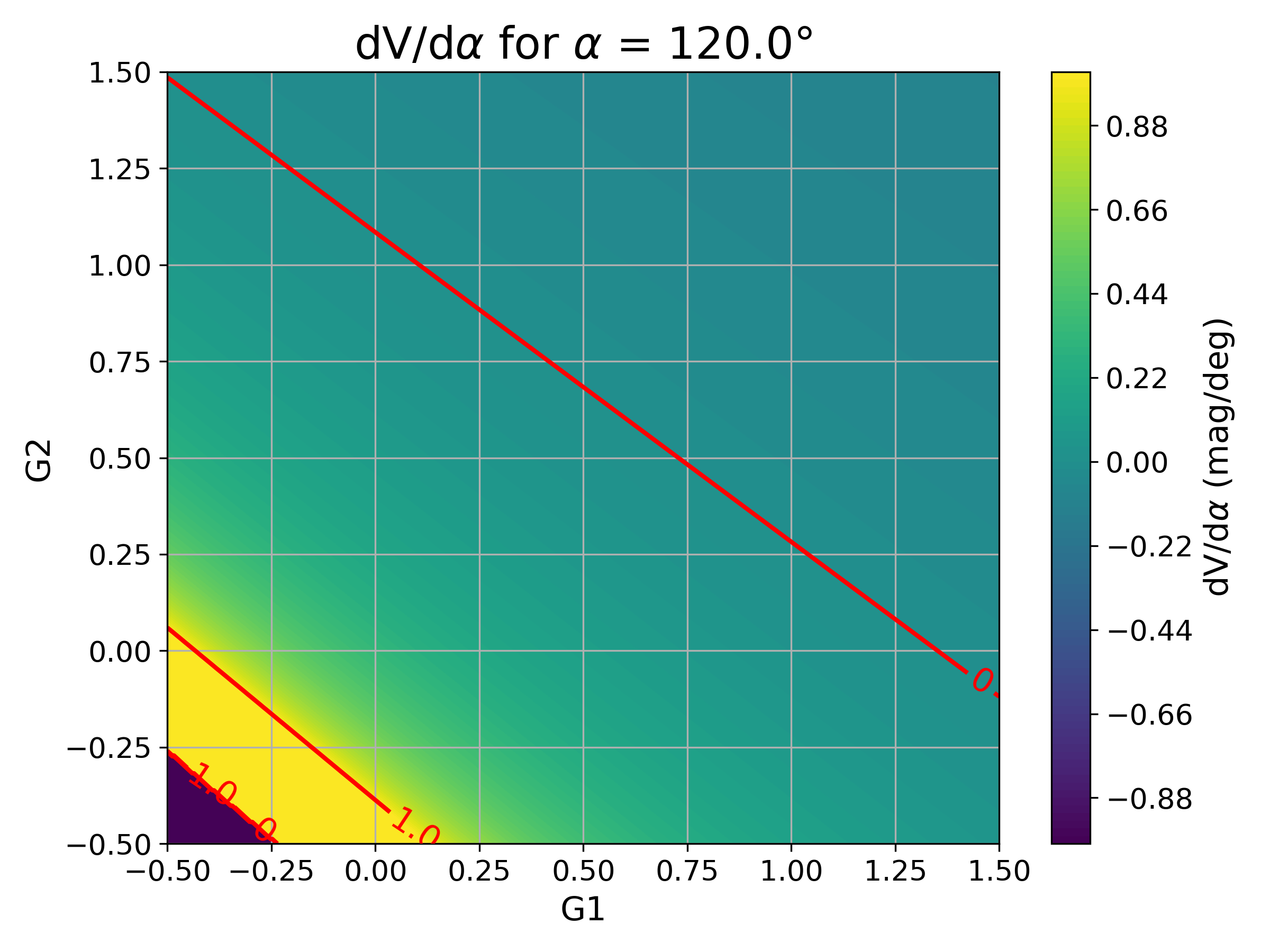}
\end{subfigure}
\hfill
\begin{subfigure}[b]{0.29\textwidth}
\includegraphics[width=\linewidth]{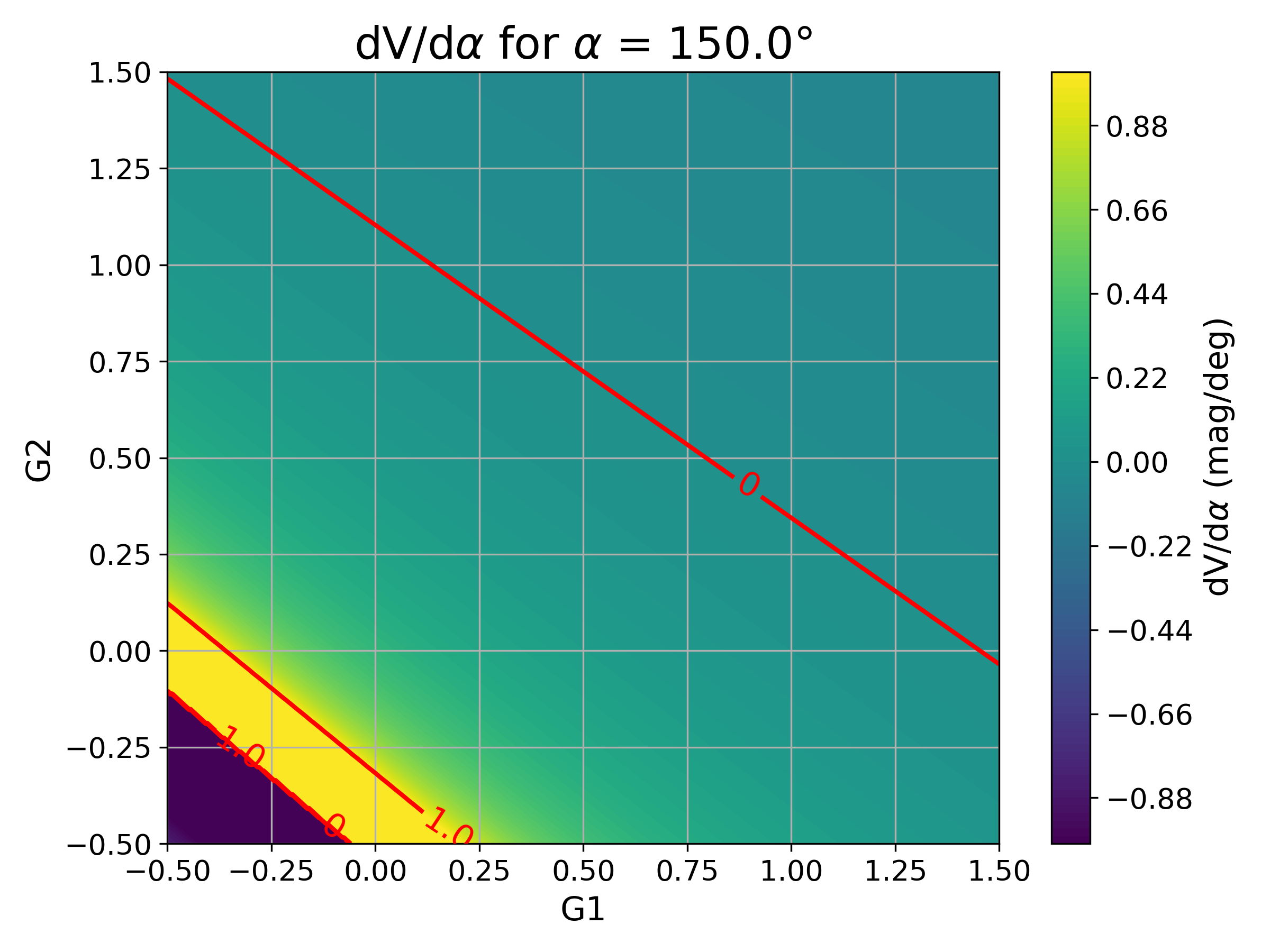}
\end{subfigure}

\caption{Heatmap showing the first derivative \( dV/d\alpha \) values for different phase angles \( \alpha \) and parameters \( G_1 \), \( G_2 \). The red lines indicate contour lines where the derivative is equal to $0.0$ mag/deg and $1.0$ mag/deg. Derivatives clipped to $-1.0$ mag/deg to $1.0$ mag/deg.}
\label{dVda}
\end{figure}
\FloatBarrier

%%%%%%%%%%--------

\subsection{Heatmaps of $\frac{dV}{d\alpha}$ for the $H,G_{12}$ phase function}

\begin{center}
    \includegraphics[width=0.5\linewidth]{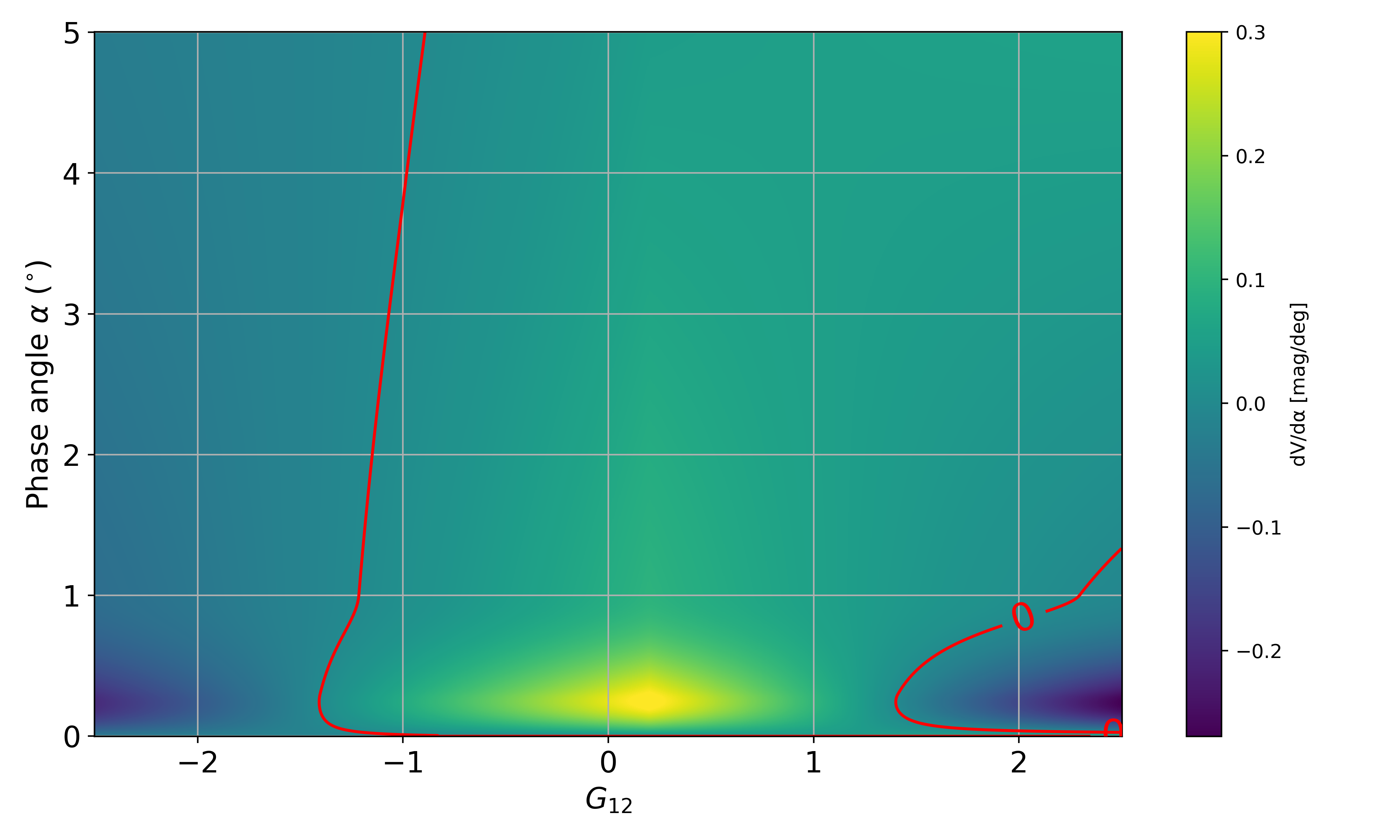}
    \includegraphics[width=0.5\linewidth]{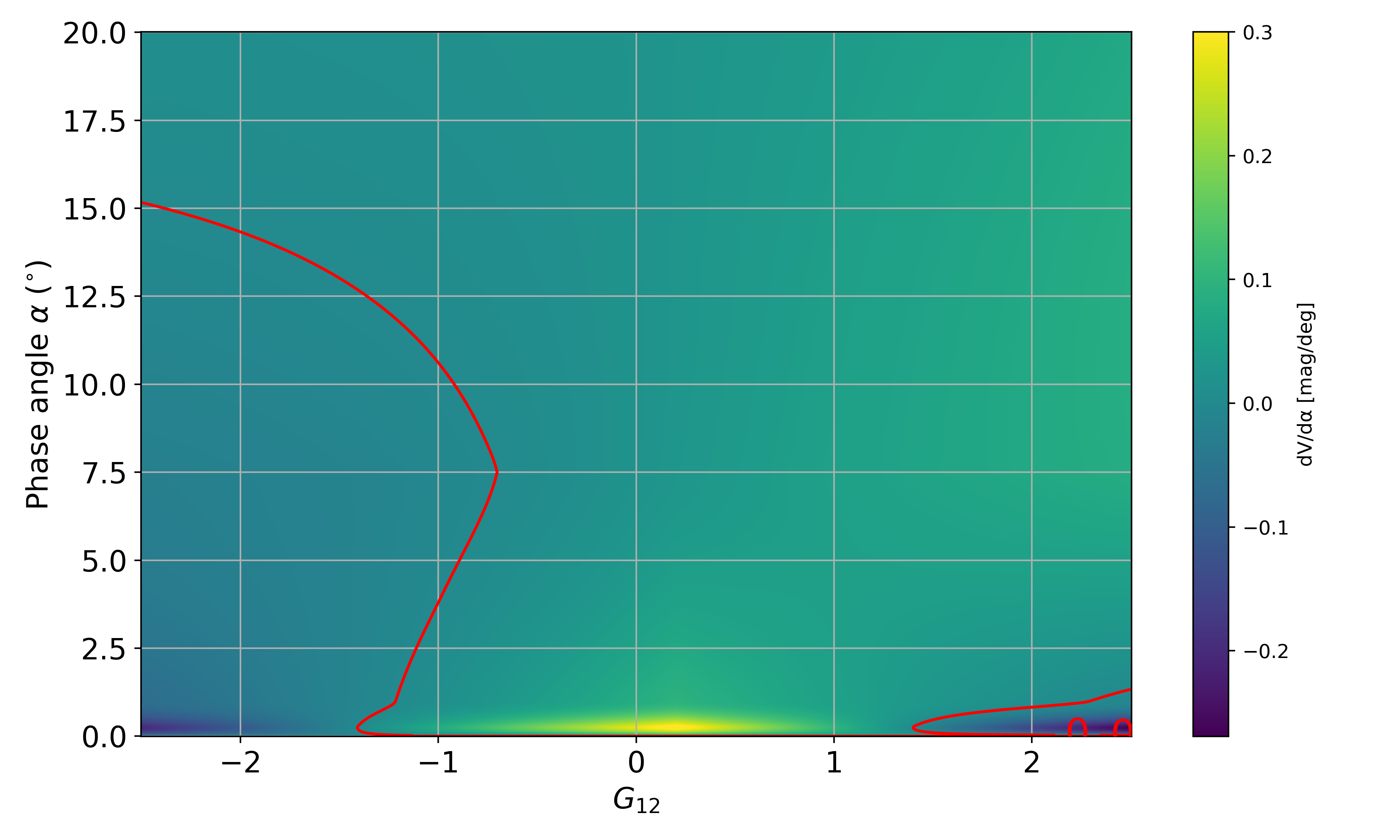}
    \includegraphics[width=0.5\linewidth]{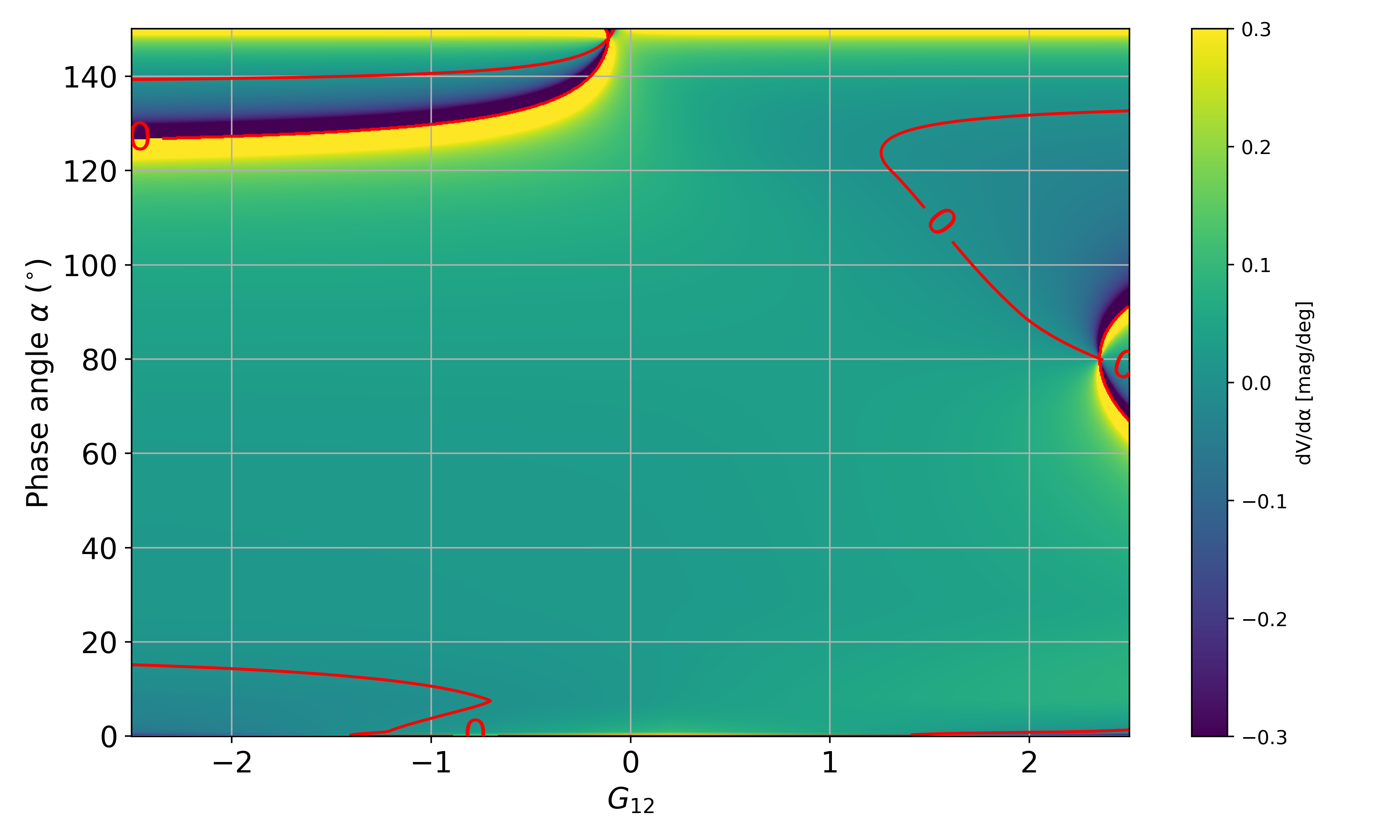}

    \captionsetup{type=figure}
    \captionof{figure}{Partial derivative of the $H, G_{12}$ phase function with respect to phase angle $\alpha$, computed for a range of $G_{12}$ values. The red contour marks the level where the derivative is zero. Derivatives clipped to $-0.3$ mag/deg to $0.3$ mag/deg.}
    \label{fig:HG12_derivative}
\end{center}

\subsection{Heatmaps of $\frac{dV}{d\alpha}$ for the $H,G^*_{12}$ phase function}

\begin{center}
    \includegraphics[width=0.5\linewidth]{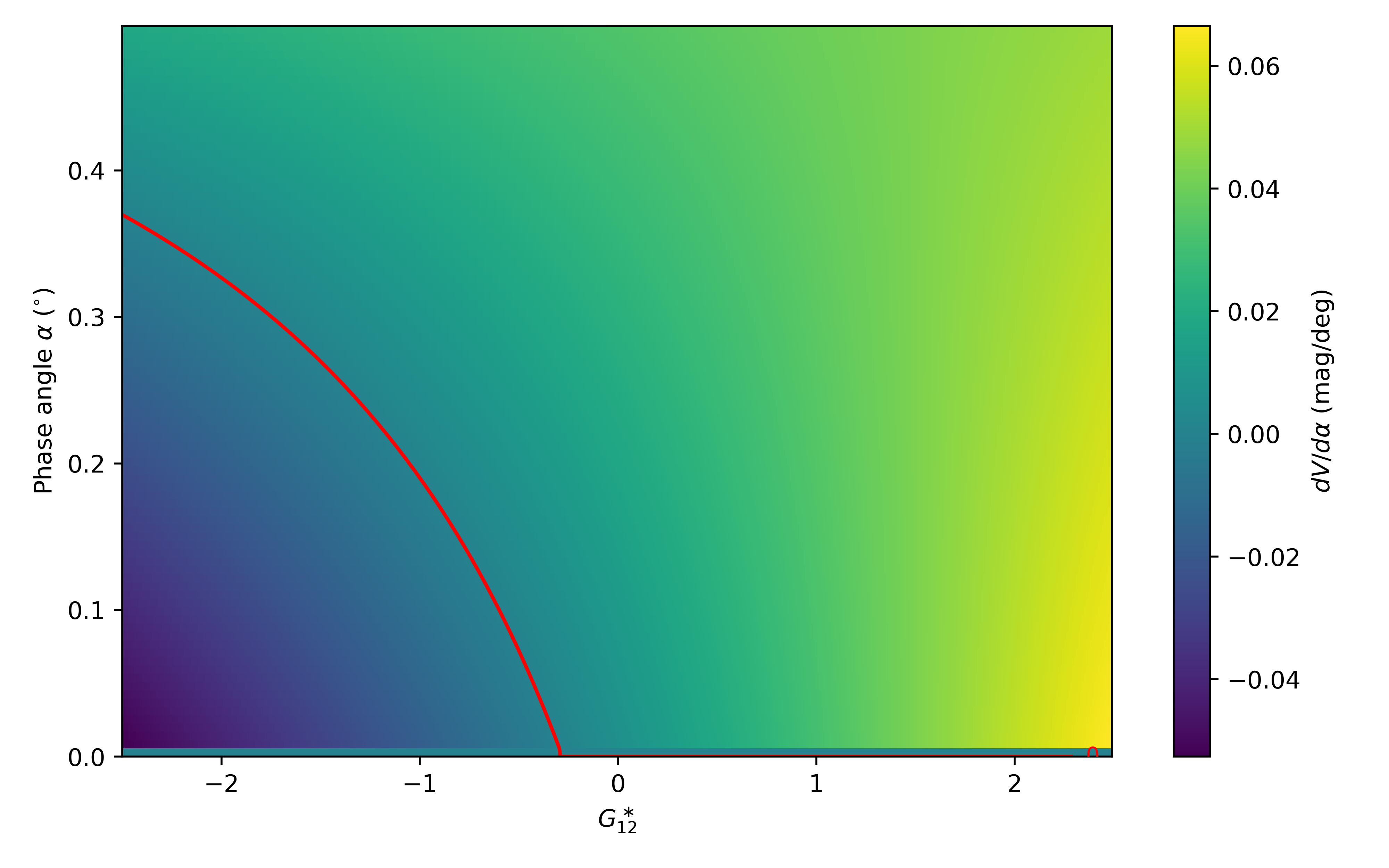}
    \includegraphics[width=0.5\linewidth]{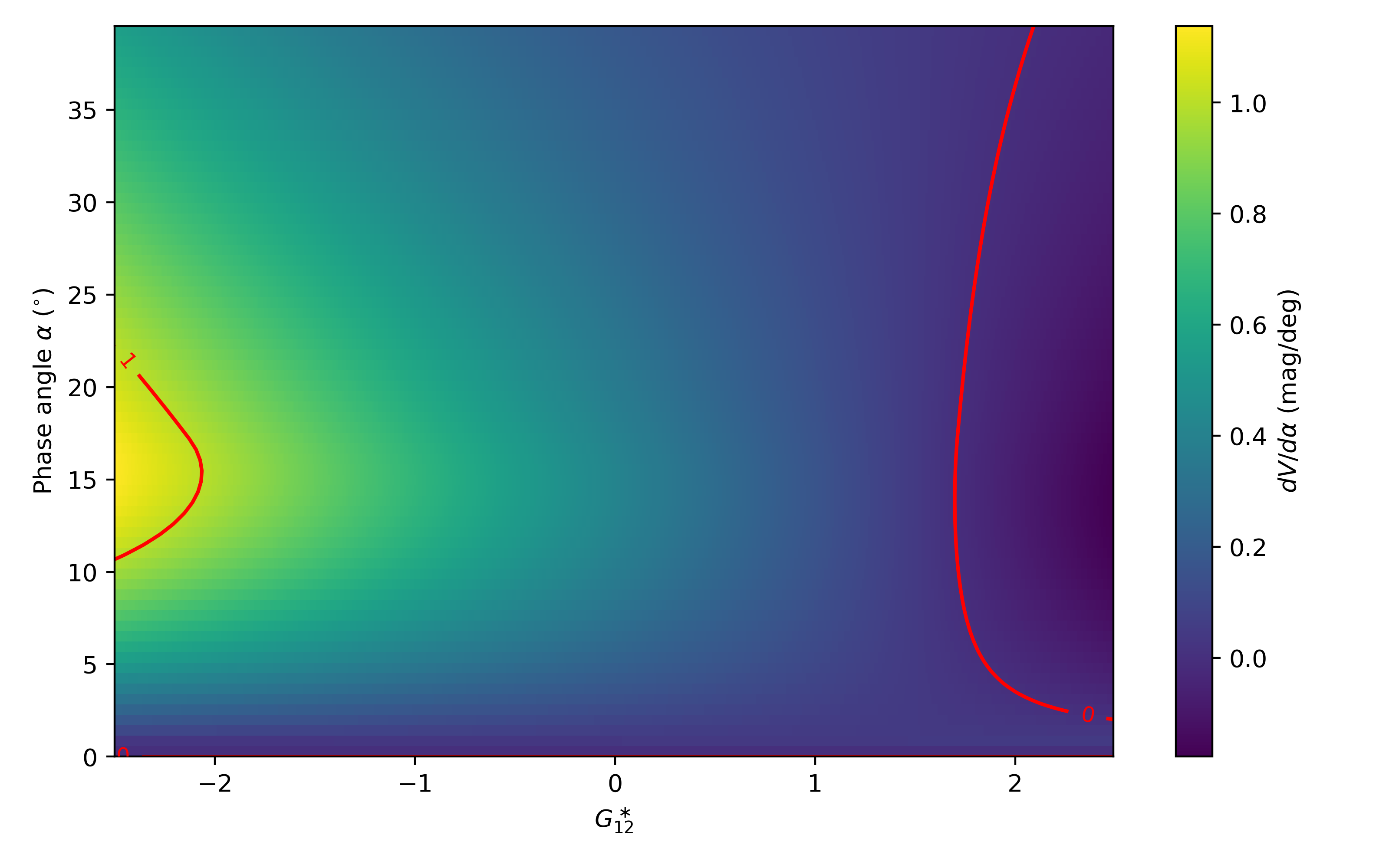}
    \includegraphics[width=0.5\linewidth]{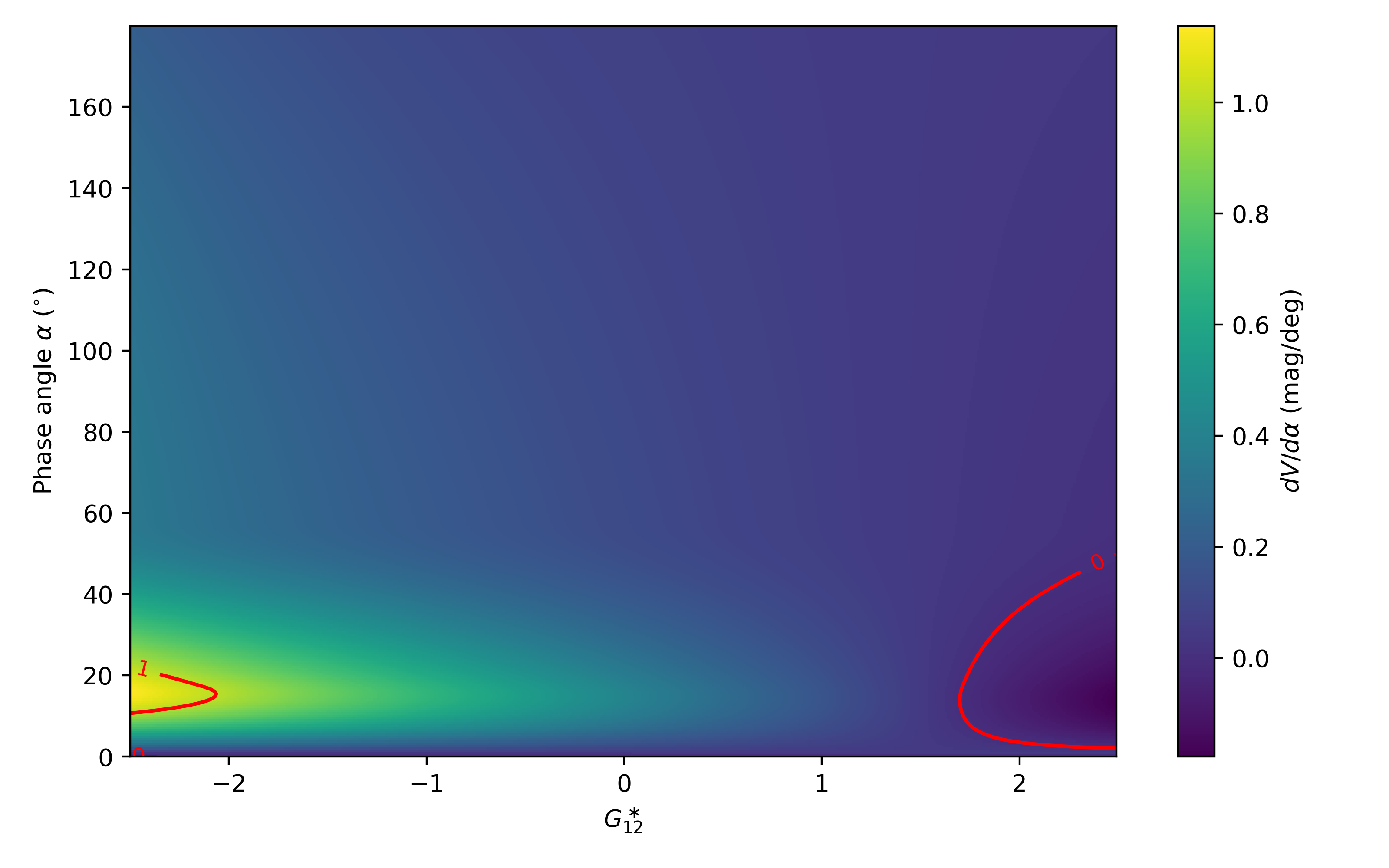}

    \captionsetup{type=figure}
    \captionof{figure}{Partial derivative of the $H, G^*_{12}$ phase function with respect to phase angle $\alpha$, computed for a range of $G^*_{12}$ values. The red contour marks the level where the derivative is zero or one.}
    \label{G12star}
\end{center}

\end{document}